\documentclass[sigconf]{acmart}
\usepackage{algorithm}
\usepackage{subcaption}
\usepackage{algpseudocode}
\usepackage{float}
\usepackage{multirow}
\usepackage{tcolorbox}
\usepackage{adjustbox}
\usepackage{enumitem}
\usepackage{listings}
\AtBeginDocument{%
  \providecommand\BibTeX{{%
    \normalfont B\kern-0.5em{\scshape i\kern-0.25em b}\kern-0.8em\TeX}}}

\setcopyright{acmcopyright}
\copyrightyear{2023} 
\acmYear{2023} 
\setcopyright{acmlicensed}\acmConference[SC '23]{The International Conference for High Performance Computing, Networking, Storage and Analysis}{November 12--17, 2023}{Denver, CO, USA}
\acmBooktitle{The International Conference for High Performance Computing, Networking, Storage and Analysis (SC '23), November 12--17, 2023, Denver, CO, USA}
\acmPrice{15.00}
\acmDOI{10.1145/3581784.3607108}
\acmISBN{979-8-4007-0109-2/23/11}

\acmConference{}{}{}

\renewcommand\footnotetextcopyrightpermission[1]{}

\setcopyright{none}
\begin{document}
\newcommand\benchname{\textit{LBench}~}
\newcommand\todo[1]{\textcolor{red}{#1}}
\newcommand\note[1]{\textcolor{blue}{#1}}
%%
%% The "title" command has an optional parameter,
%% allowing the author to define a "short title" to be used in page headers.
\title[A Quantitative Approach for Adopting Disaggregated Memory in HPC Systems]{A Quantitative Approach for Adopting Disaggregated\\ Memory in HPC Systems}
\thanks{Pre-print submitted for publication.}
\author{Jacob Wahlgren}
\email{jacobwah@kth.se}
\affiliation{%
  \institution{KTH Royal Institute of Technology}
  \country{Sweden}
}
\author{Gabin Schieffer}
\email{gabins@kth.se}
\affiliation{%
  \institution{KTH Royal Institute of Technology}
  \country{Sweden}
}
\author{Maya Gokhale}
\email{gokhale2@llnl.gov}
\affiliation{%
  \institution{Lawrence Livermore National Laboratory}
  \country{USA}
}
\author{Ivy Peng}
\email{ipeng@acm.org}
\affiliation{%
  \institution{KTH Royal Institute of Technology}
  \country{Sweden}
}

%% of authors' names for this purpose.
%\renewcommand{\shortauthors}{Anonymous}
%\settopmatter{printacmref=false}
%%
%% The abstract is a short summary of the work to be presented in the
%% article.
\begin{abstract}
Memory disaggregation has recently been adopted in data centers to improve resource utilization, motivated by cost and sustainability. %high hardware and energy costs. 
Recent studies on large-scale HPC facilities have also highlighted memory underutilization. 
A promising and non-disruptive option for memory disaggregation is rack-scale memory pooling, where node-local memory is supplemented by shared memory pools. %Previous works have extended the roofline model to search for HPC system designs using disaggregated memory.
This work outlines the prospects and requirements for adoption and clarifies several misconceptions. %using a workload-driven quantitative approach. 
We propose a quantitative method for dissecting application requirements on the  memory system from the top down in three levels, moving from general, to multi-tier memory systems, and then to memory pooling. 
%First, general workload characteristics include classic roofline modeling and access patterns and prefetching analysis. Second, for multi-tier memory the access distribution across local and remote memory is analyzed. 
We provide a multi-level profiling tool and \benchname to facilitate the quantitative approach.  
%study memory interference from pool sharing.
%-- a key challenge for memory disaggregation.
%... that combines an adapted roofline model and application profiles of memory access patterns to identify optimization opportunities and hardware limitations. We design and develop a set of benchmarks, profilers, and metrics for constructing a roofline model and application profiles.
We evaluate a set of representative HPC workloads on an emulated platform.
%Based on the results, we present guidelines for deploying and optimizing HPC applications on disaggregated systems.
Our results show that prefetching activities can significantly influence memory traffic profiles. 
Interference in memory pooling has varied impacts on applications, depending on their access ratios to memory tiers and arithmetic intensities.
%We show in three use cases that with guidance from application and system profiles, the performance impact of disaggregated memory can be mitigated for both interference sensitive and insensitive applications by adapting the deployment configuration and data placement.
Finally, in two case studies, we show the benefits of our findings at the application and system levels, achieving 50\% reduction in remote access and 13\% speedup in BFS, and reducing performance variation of co-located workloads in interference-aware job scheduling. 
%We identify poor data placement in one of the workloads and improve performance by up to 13\% by reducing remote memory accesses. We also show how interference-aware job scheduling can improve performance by co-locating compatible workloads only.
\end{abstract}

%%
%% The code below is generated by the tool at http://dl.acm.org/ccs.cfm.
%% Please copy and paste the code instead of the example below.
%%
\begin{CCSXML}
<ccs2012>
   <concept>
       <concept_id>10010583.10010786.10010809</concept_id>
       <concept_desc>Hardware~Memory and dense storage</concept_desc>
       <concept_significance>500</concept_significance>
       </concept>
   <concept>
       <concept_id>10010520.10010521.10010542.10010546</concept_id>
       <concept_desc>Computer systems organization~Heterogeneous (hybrid) systems</concept_desc>
       <concept_significance>500</concept_significance>
       </concept>
   <concept>
       <concept_id>10010583.10010786.10010808</concept_id>
       <concept_desc>Hardware~Emerging interfaces</concept_desc>
       <concept_significance>500</concept_significance>
       </concept>
 </ccs2012>
\end{CCSXML}

\ccsdesc[500]{Hardware~Memory and dense storage}
\ccsdesc[500]{Computer systems organization~Heterogeneous (hybrid) systems}
\ccsdesc[500]{Hardware~Emerging interfaces}

%%
%% Keywords. Separate the keywords with commas.
\keywords{HPC system, disaggregated memory, multi-tier memory}

\maketitle

\section{Introduction}
\begin{figure}[bt]
\centering
\includegraphics[height=2.7cm,width=\linewidth]{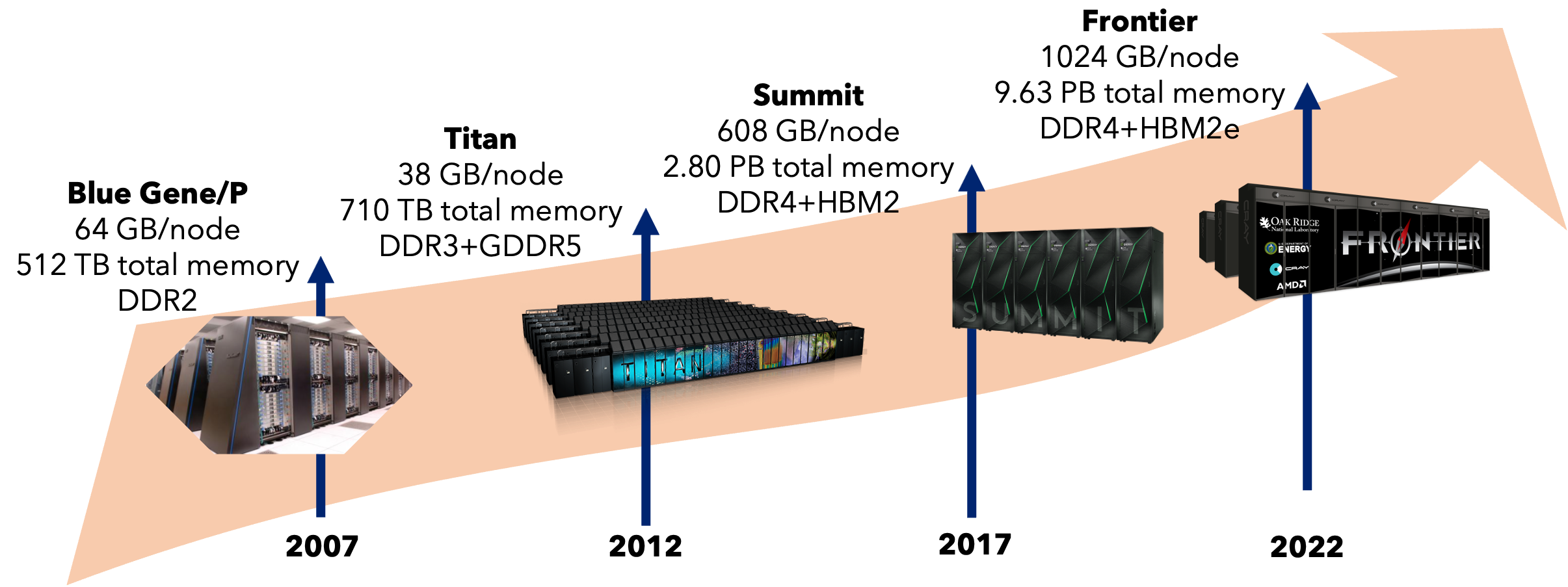}
\caption{The evolution of memory characteristics of top leadership supercomputers in the past 15 years.}
\label{fig:timeline}
\end{figure}

The memory subsystem is a major consideration for cost and performance of large-scale clusters.
%In the last decade, the divergence between compute and memory continues deepening.
In the last decade, the memory bandwidth wall has emerged as the bandwidth per core continues to decrease. To satisfy the increasing demand, memory capacity and bandwidth per node have increased dramatically in the last 15 years, as illustrated in Figure~\ref{fig:timeline}. New memory technologies like high-bandwidth memory (HBM) can provide significant bandwidth. However, the cost per bit is three to five times that of regular DDR due to a complex production process and high market demand~\cite{memoryCost2023}. Hence, a hybrid of DDR and HBM has become the de-facto memory configuration on today's HPC systems as we show in Table~\ref{tab:memory}. For instance, the No. 1 supercomputer, Frontier, features 512 GB DDR4 and 512 GB HBM per compute node. 

The node architecture in clusters and data centers today is monolithic -- compute and memory resources are tightly coupled within a node's boundary. In a cloud environment, multiple jobs may share one server through virtualization, and still, recent works find that substantial node memory is not utilized~\cite{pond23}. In an HPC environment, the resource allocation is even more coarse-grained because jobs typically do not share a node. Combined with high memory capacity per node, this means that memory is often an under-utilized resource. Recent studies on leadership facilities have found that fewer than 15\% jobs use more than 75\% of the node memory, and that 50\% of the time less than 12\% of the total memory is in use~\cite{peng2020memory,lbl_dm,peng2021holistic, michelogiannakis2022case}.

Recently, major cloud operators have started exploring memory disaggregation to address the challenge of memory utilization, including Meta~\cite{maruf2023tpp} and Microsoft Azure~\cite{pond23}. At a high level, the idea is to have a portion of memory resources provided on-demand through remote memory pooling. In memory pool-based systems, each node has a fixed local memory and a variable amount of fabric-attached remote memory, effectively a form of a multi-tier memory system~\cite{ewais_disaggregated_2023}. Previous works have explored using node-local persistent memory, like Intel Optane DC PM, as a second memory tier~\cite{google_pm,peng2020demystifying}. Since Optane is discontinued, and with the recent advances in cache-coherent interconnect protocols, i.e., Compute Express Link (CXL) type 3 devices, memory pooling becomes an emerging option for implementing the second tier~\cite{tsai2020disaggregating}.

Existing works in multi-tier systems with memory pooling focus on cloud infrastructures~\cite{lim2009disaggregated,lim2012system,gu2017efficient,pinto2020thymesisflow,Gouk2022,maruf2023tpp,pond23}. Cloud providers are motivated to guarantee a certain Quality of Service (QoS) while minimizing their cost, e.g. through increased system utilization. While their findings and solutions are useful for informing HPC systems, HPC-oriented studies are needed for factoring in differences in the deployment environment, user expertise, and expectation. Therefore, this work aims to lay out the general and practical considerations for adopting disaggregated memory on HPC systems.

Although disaggregated memory is a type of multi-tier memory, it differs from traditional Non-Uniform Memory Access (NUMA) systems, which are symmetric, with each NUMA domain containing the same compute and memory resources, and optimizations focus on moving computing to cores in the same domain as the accessed memory pages. Instead, a tiered memory system could be asymmetric, where some tiers are CPU-less, and optimizations focus on placing hot pages in faster tiers~\cite{pond23}.

% our method
We propose a three-level quantitative approach for dissecting application requirements on the memory system from the top down. At the top level, an application's intrinsic requirements on the memory subsystem are captured and these properties remain preserved across different memory systems. The second level quantifies the impact of a general multi-tier memory system and captures the memory access ratios across different tiers. Finally, the third level quantifies the impact of memory interference, a specific challenge of pooling-based multi-tier memory.

Several existing works focus on understanding the performance degradation or optimizing the data placement on disaggregated memory~\cite{pinto2020thymesisflow,Gouk2022,wahlgren2022evaluating,maruf2023tpp,pond23}. However, instead of providing yet another optimization work of data placement, this paper builds a generic framework to study an application's memory behavior and identify optimization priority and limits, as well as influence deployment options. To facilitate the quantitative study, we provide a multi-level profiling tool and also \benchname, a benchmark for quantifying the impact of memory interference. Our methodology is applied in a set of diverse applications, including NekRS, SuperLU, Hypre, HPL, BFS, and XSBench. Finally, in two case studies, we show that findings from our approach can guide application-level optimizations on data placement, reducing remote access by 50\% and improving the performance of BFS by 13\%, as well as guide system-level scheduling to reduce performance variation of co-located jobs. 
%improve the overall efficiency of co-located jobs, improving the performance of Hypre by 4\%.
% our results
%Existing works on analyzing performance degradation on disaggregated memory~\cite{mchpc_paper}. The main challenge of adopting memory pooling in HPC systems come from potential performance degradation. therefore, most previous works~\cite{meta_paper,mchpc_paper,xx} focus on understanding the performance impact on application when different memory capacity is backed by the remote memory. In contrary, we believe that integration of disaggregated memory into HPC systems increases the dimension of deployment decisions on HPC system but not necessarily means performance impact. In fact, our extended roofline model indicates an increased theoretical memory bandwidth when moving from monotholic to disaggregated architecture. Therefore, our work provides a methodology for incorporating the accurate memory requirements on disaggregated memory into user's deployment decisions.  
% our contributions
%\vfill\noindent

We summarize our contributions in this work as follows:
\begin{itemize}[leftmargin=*]
\item We describe the prospects and requirements for adopting disaggregated memory in HPC systems.
\item We propose a quantitative method for dissecting application requirements on memory systems in three levels, from general to multi-tier memory systems and memory pooling.
%\item We propose an extended roofline model for identifying performance bottlenecks and optimization opportunities.
\item We develop a multi-level profiler for facilitating the quantitative method and \benchname for quantifying interference on memory pooling.
\item We evaluate our method in NekRS, SuperLU, Hypre, HPL, BFS, and XSBench on multiple emulated system configurations and identified key insights.
\item We demonstrate the usage of our method in two case studies, for optimizing the memory access ratios at the application level and interference-aware job scheduling at the system level.
\end{itemize}

\section{Motivation and Background}\label{sec:bg}
\begin{table}[bt]
\begin{adjustbox}{width=\linewidth,center}{
\centering
\begin{tabular}{|c|c|c|c|c|c|c|c|}
 \hline
 \textbf{Rank} & \textbf{DDR/node} &\textbf{HBM/node} &\textbf{HBM BW/node}&\textbf{Nodes} &\textbf{Est. DDR cost} &\textbf{Est. HBM cost} \\\hline\hline
%&\textbf{Memory Power}
Frontier~\cite{frontierSpecs} 
    & 512~GB & 512~GB & 12.8~TB/s & 9,408 & \$~34~M & \$~135~M \\\hline
    % DDR4:4.8~PB, HBM2e:4.8~PB
Fugaku~\cite{fugakuSpecs}
    & -- & 32~GB & 1.0~TB/s & 158,976 & -- & \$~142~M \\\hline
    % HBM2:5.1~PB
LUMI-G~\cite{lumiSpecs}
    & 512~GB & 512~GB & 12.8~TB/s & 2,560 & \$~9.2~M & \$~35~M\\\hline
    % DDR4:1.3~PB, HBM2e:1.3~PB
Leonardo~\cite{leonardoSpecs}
    & 512~GB & 256~GB & 8.2~TB/s & 3,456 & \$~12~M & \$~25~M \\\hline
    % DDR4:1.8~PB, HBM2:0.88~PB, BW assuming A100 DGX (not custom)
Summit~\cite{summitSpecs}
    & 512~GB & 96~GB & 5.4~TB/s & 4,608 & \$~17~M & \$~12~M \\\hline
    % DDR4:2.4~PB, HBM2:0.44~PB
Sierra~\cite{sierraSpecs}
    & 256~GB & 64~GB & 3.6~TB/s & 4,284 & \$~7.7~M & \$~7.7~M \\\hline
    % DDR4:1.1~PB, HBM2:0.27~PB
Sunway~\cite{sunwaySpecsReport}
    & 32~GB & -- & -- & 40,960 & \$~9.2~M& -- \\\hline
    % DDR3:1.3~PB
Perlmutter (GPU)~\cite{perlmutterSpecs}
    & 256~GB & 160~GB & 6.2~TB/s & 1,536 & \$~2.8~M & \$~7.0~M \\\hline
    % DDR4:0.39~PB, HBM2e:0.25~PB
Selene~\cite{seleneSpecs}
    & 1~TB & 640~GB & 16~TB/s & 280 & \$~2~M & \$~4.9~M \\\hline
    % DDR4:0.28~PB, HBM2e:0.18~PB
Tianhe-2A~\cite{tianheSpecs}
    %& 64~GB & 128~GB (DDR4) & -- & 16,000 & \$~7.2M & \$~49M \\\hline    
    & 192~GB & -- & -- & 16,000 & \$~21.6~M & -- \\\hline
    %  DDR3:1.0~PB, DDR4:2.0~PB
\end{tabular}
}\end{adjustbox}
% Bandwidth is only for HBM
\caption{A summary of memory configuration on Top 10 supercomputers and the estimated memory cost based on HBM having 3$\times$-5$\times$ unit price of DDR~\cite{memoryCost2023}.}%estimated at 7\$/GB for DDR, 8\$/GB for HBM)
\label{tab:memory}
\end{table}

{\iffalse
DRAM
[1] https://aiimpacts.org/trends-in-dram-price-per-gigabyte/ => interesting, but stops in 2018, consumer grade
[2] John McCallum (source for [1]), continues to 2022: https://jcmit.net/memoryprice.htm
[3] https://memory.net/memory-prices/ 
    Interesting, cf Excel file for aggregated results.
    The web archive version (2018-12) aligns with the other source
    => USD 7/GB seems reasonable for server grade memory

HBM2
[1] https://www.gamersnexus.net/guides/3032-vega-56-cost-of-hbm2-and-necessity-to-use-it
    In 2017: \$150 for 8GB => \$19/GB
[2] https://semiengineering.com/whats-next-for-high-bandwidth-memory/
    In 2016: \$120/GB
    In 2019: \$120 for 16GB => \$7.5/GB
\fi}

%high cost, low utilization, diverse workload, scalable upgrade
In the past 15 years, as workloads on HPC systems continue evolving, high computation throughput, massive datasets, complex workflow, and emerging machine learning components drive up the requirements on the memory system (see Figure~\ref{fig:timeline}). Table~\ref{tab:memory} summarizes the memory configuration on today's top 10 supercomputers~\cite{top500}, where memory is becoming a major cost factor.
%As impressive as the amount of memory resources per node is, the high cost estimated from DDR4's \$ per bit and the relative price of HBM to DDR, which is $3-6\times$ in the past few years due to the high demand from the market.
%\cite{nersc20}
Meanwhile, studies on multiple HPC facilities, including NERSC's Cori supercomputer~\cite{michelogiannakis2022case} and Livermore Computing's clusters~\cite{peng2020memory}, have shown that the utilization of memory resources can be as low as only 15\% of scientific workloads utilizing at least 75\% memory resources. In data centers, under-utilization of memory has motivated major cloud providers to leverage disaggregated memory to provide memory resources as needed and improve overall utilization~\cite{gu2017efficient,maruf2023tpp,pond23}.

%definition of disaggregated memory, two possible architectures
\textit{Disaggregated Memory} refers to a type of architecture that decouples memory resources from compute resources so that a flexible amount of memory resources can be provisioned to each workload. In contrast, all the supercomputers in Table~\ref{tab:memory} have a fixed amount of memory in DDR and HBM, so a job cannot use more than the specified memory capacity on a node, or it will face out-of-memory (OOM) errors and abort.

\begin{figure}
\centering
\includegraphics[width=\columnwidth]{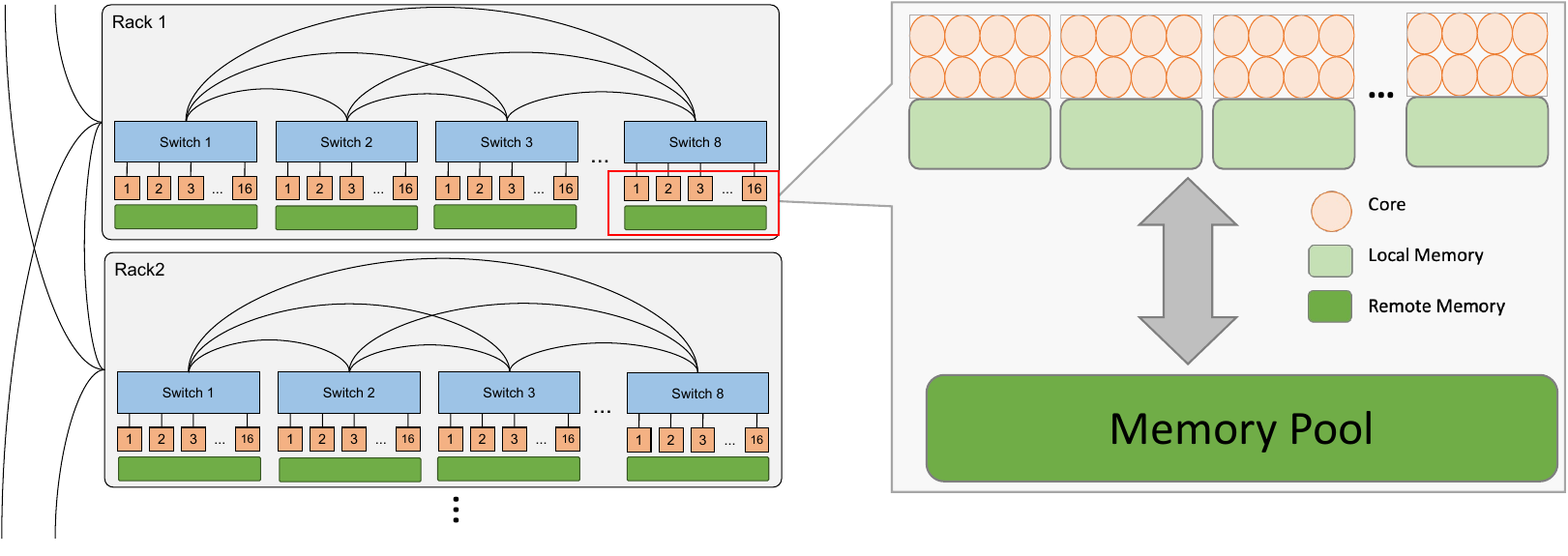}
\caption{An HPC system architecture with rack-scale memory disaggregation. Each node has a fixed node-local memory. Nodes in the same rack also share a memory pool.}
\label{fig:arch}
\end{figure}

There are two categories of disaggregated memory architectures -- split and pool~\cite{ewais_disaggregated_2023}. In a split architecture, nodes can borrow memory from each other in a peer-to-peer fashion. In contrast, a \textit{Memory Pool} provides a dedicated memory resource shared by multiple nodes. Memory disaggregation enables peak-of-sums provisioning rather than sum-of-peaks provisioning, reducing the required resources on a system level~\cite{lim2009disaggregated}. Hence, the total ownership cost (TCO) can be reduced, resulting in a great incentive for major cloud providers. Previous network-based disaggregation solutions face challenges of performance degradation. The recent development of high-performance cache-coherent interconnects, like the CXL standard, dramatically improves the feasibility and is endorsed by all major vendors. For HPC systems, a rack-scale memory pooling architecture, as illustrated in Figure~\ref{fig:arch}, may be the most feasible and user-transparent design in the near term~\cite{ewais_disaggregated_2023,matsuoka2023myths}. This design bears similarity with the LLNL's HPE Rabbit storage design~\cite{llnlrabbit21}, where compute blades in a rack share a pool of SSD storage resources.

\subsection{Challenges and Misconceptions}\label{sec:challenges}
A disaggregated memory system is a specific implementation of \textit{Muti-tier Memory} systems. Muti-tier memory is composed of different tiers with distinctive performance and capacity. For instance, 8 out of the top 10 supercomputers in Table~\ref{tab:memory} use HBM-DDR-based multi-tier memory systems. For the architecture presented in Figure~\ref{fig:arch}, the node-local memory (light green) forms a top tier, and the memory pool (dark green) forms the bottom tier of a node's memory system.

\textit{Memory Interference} in a memory pool refers to the influence from other nodes sharing the same pool. Through the shared memory pool, interference can cause unpredictable performance degradation of one job due to jobs on other nodes. For example, one cause for the performance variation is unrelated jobs on different nodes competing for the link bandwidth between compute node and memory pools.

\textit{Misconceptions} of multi-tier memory include that the memory bandwidth is lower than homogeneous memory. In fact, from the hardware perspective, adding additional memory tiers (channels) can increase the aggregate bandwidth. However, a challenge is to utilize it effectively. Another misconception is that application performance will always be degraded.
%This may be true for data center workloads.
Distributed-memory HPC applications have the option to minimize exposure to the pooled memory tier by scaling to more compute nodes instead.

%\textbf{Data Movement} the Unique Challenge is a composable architecture instead of a fixed one, where the interconnect bandwidth may be a variable. Two levels of granularity are commonly used -- page level and cache line. They are not mutual exclusive. Several existing works propose a hybrid of the two. \todo{Common Optimizations}--> instead, this work provides a performance model that trades off the page migration (and resulted local memory access) and remote cache line access.

\subsection{Practical Considerations for Adoption}\label{sec:adoption}
The differences between HPC and cloud systems have to be considered for adoption. HPC applications consist of many optimized numerical kernels, run in multiple bare-metal nodes tightly coupled by MPI communication, without virtualization or node sharing. Cloud workloads like web services, and databases, are optimized for tail latency of requests. Cloud environments are virtualized and users share physical nodes. Also, the incentive structures are different. While cloud providers are motivated to reduce costs while still meeting the same service quality, HPC users are responsible for delivering the performance of their applications. Thus, approaches derived for cloud may be inapplicable on HPC, e.g., prefetching is found to be harmful in \cite{mahar2023workload}, but we find it necessary for HPC. Cloud solutions are mostly based on extended hypervisors or managed runtimes, while HPC infrastructures do not have such layers. Here, we highlight several practical requirements for adopting disaggregated memory in HPC systems.

\textbf{Low Porting Efforts}. For wide adoption in HPC, the changes in architecture should aim to require low porting efforts. One solution for transparent porting is extending current NUMA system support. As an example, a recent kernel patch implements non-uniform interleaving policies~\cite{weiner_2022} which enables applications to transparently utilize the aggregated bandwidth of tiered memory~\cite{sun2023demystifying}. Alternatively, existing memory allocators can be extended to support multi-tier systems.

\textbf{Low Performance Variation}. Many runtime solutions for optimizing data placement on multi-tier memory systems are proposed. While automatic NUMA support requires low porting efforts, runtimes take time to collect enough information for decisions and are often slow in adapting to changes in access patterns, resulting in performance variation from run to run. HPC applications, however, demand more deterministic and reproducible performance.

\textbf{Incentive for Computing Facility}. End users of HPC systems are usually not incentivized to improve system-level resource utilization. Therefore, the computing facility needs to be the main driver for adoption. Potential incentives for facilities to adopt memory pooling include sustainability, the feasibility of separate upgrading of system components, and reduced costs. 

%As explained by the main rationale of using as many nodes as necessary to fit into the fast tier memory and leaving the lower tier underutilized. 
%possible optimization to supply hints to schedulers, references to Adrias and Google
%Previous deployment of node-local NVMe for data staging is not widely adopted as it still requires user explicit changes in mounting a new filesystem . 

\section{Methodology}
\begin{figure}
\centering
\includegraphics[width=\linewidth]{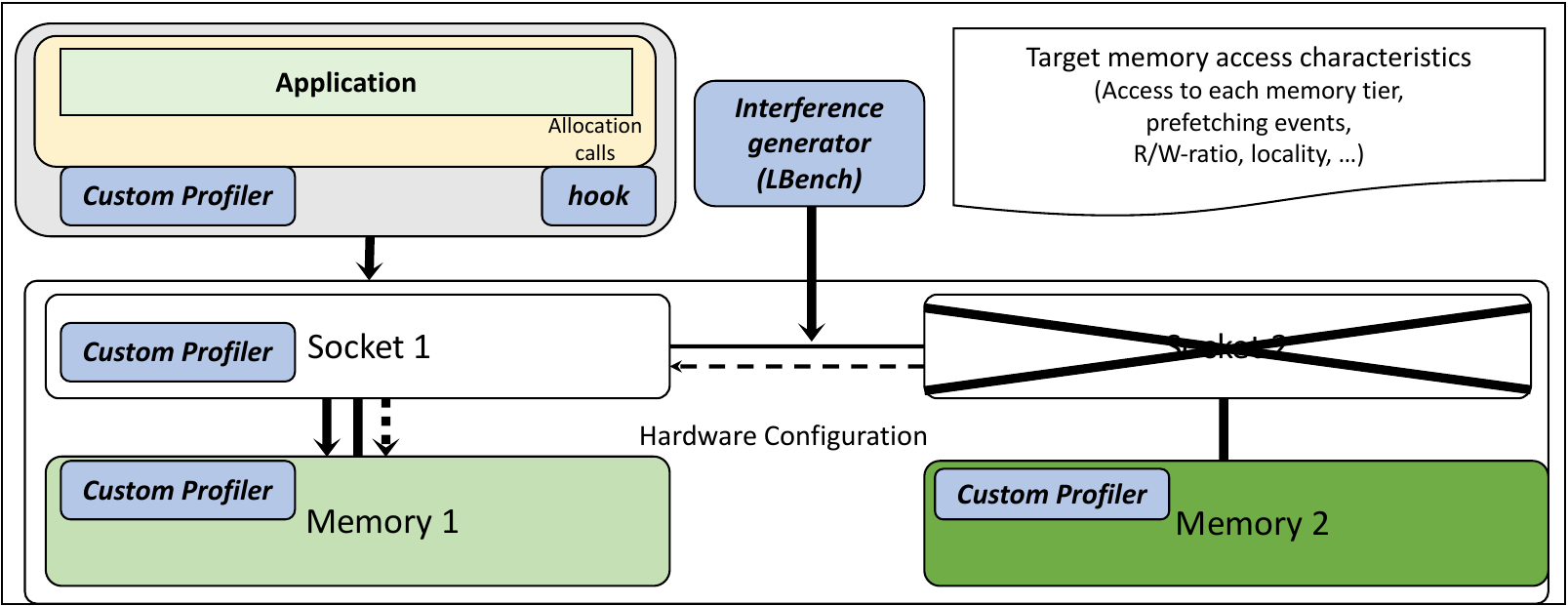}
\caption{The overall architecture of the three-level memory-centric profiler and the emulation platform.}
\label{fig:profiler}
\end{figure}

\begin{figure}
    \centering
    \includegraphics[width=0.9\linewidth]{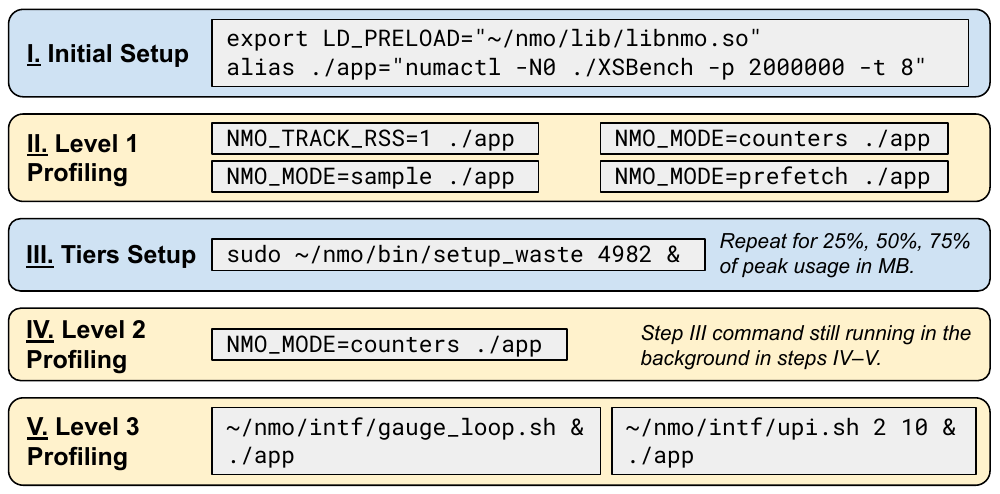}
    \caption{A step-by-step overview of the profiling workflow with example commands.}
    \label{fig:workflow}
\end{figure}

In this work, we employ a three-level top-down approach. In the first level, we identify an application's intrinsic requirements on memory systems, independent of exact system architecture. The second level extends the application requirements onto general multi-tier memory systems and leverages a memory roofline model. The final level investigates the specific memory interference challenge when the lower tier is backed by memory pooling on an HPC architecture illustrated in Figure~\ref{fig:arch}. Figure~\ref{fig:workflow} presents the experimental workflow, where each level of execution collects profiling information that can be visualized separately.

\subsection{Multi-level Profiling}
We develop a multi-level profiler to support the three-level memory-centric analysis methodology. The profiler uses low-overhead hardware performance counters to capture application-level metrics, such as memory accesses in different levels of the memory hierarchy. Besides program-level profiling, the profiler also offers APIs, i.e., \verb|pf_start("tag")| and \verb|pf_stop()|, for simple tracing support to attribute results to specific kernels. This section details our methodology based on Linux and the Intel Skylake-X architecture. Key performance events are also available on other architectures, e.g. AMD's Instruction-Based Sampling and ARM’s Statistical Profiling Extension. Kernel-based page-level profiling may be used as an alternative, but would require extensive changes. Our method can be adapted for general multi-tier memory (e.g. DDR+HBM, PM+DDR), only needing to change selected performance events.

\textbf{Level 1: General Characteristics.}
The first level of profiling aims to understand an application's requirements on the memory subsystem. These include its arithmetic intensity, memory capacity usage, bandwidth usage, and access pattern.
The arithmetic intensity is measured using hardware performance counters, enabling us to place the application into a roofline model. The number of bytes loaded is measured with the \verb|OFFCORE_RESPONSE:L3_MISS| events. The offcore events include all memory loads, including hardware prefetchers. The memory capacity usage is measured by sampling the \verb|numa_maps| file in \verb|procfs|. The memory access pattern is measured in two ways, first using precise event-based sampling (PEBS) to record the virtual address of demand load misses. Secondly, counters related to hardware prefetching are measured to understand if the access pattern is predictable.

\textbf{Level 2: Multi-Tier Memory Access.}
In a multi-tier environment, we define two key metrics. The \textit{remote capacity ratio} is the ratio of lower-tier memory to total available memory. In our setup, it can be measured from \verb|numa_maps|. The \textit{remote access ratio} is the ratio of memory accesses to a lower tier. In our setup, it is measured using the \verb|LOCAL_DRAM| and \verb|REMOTE_DRAM| offcore events. The event-based sampling of memory accesses is also extended to multiple tiers by separating cache miss events to local or remote memory.

\textbf{Level 3: Memory Interference.}
A disaggregated memory system is a specific form of multi-tier memory system. When memory pools are used to decouple memory from compute, interference from other nodes sharing the memory pool may impact performance. In Section~\ref{sec:benchmark}, we present a benchmark for measuring both the interference sensitivity and induced interference of an application. The interference sensitivity determines the performance degradation due to remote memory interference, and the induced interference determines how much interference an application causes for others. We measure the injected traffic at the system level using the UPI counters \verb|sktXtraffic| in Intel PCM. Note that link traffic may exceed the peak data bandwidth due to protocol overheads.

\begin{comment}

\begin{figure}[bt]
    \centering
    \includegraphics[height=3cm,width=\columnwidth]{figs/memory-scaling-bfs-timeline.png}
    \caption{A timeline of object-level memory access profiling on BFS.\todo{update to NekRS}}
    \label{fig:offload}
\end{figure}
\end{comment}

\subsection{Measuring Memory Interference}\label{sec:benchmark}
We developed a benchmark called \benchname based on the methodology in~\cite{delimitrou2013ibench} for injecting and quantifying interference on the link to the memory pool. The benchmark allocates an array on the memory pool and performs a simple numerical kernel similar to Empirical Roofline Toolkit~\cite{williams_roofline_2009} on the array. We define the level of interference (denoted $LoI$) as the percentage of generated link traffic compared with the peak link traffic (which is 1 flop, 12 threads on our testbed). $LoI$ is configured by varying the number of floating-point operations per array element in the kernel. We present a code snippet of the kernel's inner loop below.%More computations lead to lower memory interference, and vice versa. 
\begin{center}
\begin{minipage}{.7\linewidth}
\begin{lstlisting}
if (NFLOP % 2 == 1)
    beta = A[i] + alpha;
const int NLOOP = NFLOP/2;
#pragma GCC unroll 16
for (int k = 0; k < NLOOP; k++)
    beta = beta * A[i] + alpha;
A[i] = beta;
\end{lstlisting}
\end{minipage}
\end{center}

\begin{comment}
\begin{lstlisting}
template<const int NFLOP>
void kernel(size_t nsize, size_t ntrails, double* __restrict__ A)
{
    double alpha = 0.5;

    for (size_t j = 0; j < ntrails; j++) {
        for (size_t i = 0; i < nsize; i++) {
            double beta = 0.8;

            if (NFLOP % 2 == 1)
                beta = A[i] + alpha;

            const int NLOOP = NFLOP/2;
#pragma GCC unroll 16
            for (int k = 0; k < NLOOP; k++)
                beta = beta * A[i] + alpha;

            A[i] = beta;
        }

        alpha = alpha * (1 - 1e-8);
    }
}
\end{lstlisting}
\end{comment}

By measuring the link traffic using level 3 profiling, we determine the number of flops per element corresponding to each level of intensity (i.e., $LoI=10, 20,...$). One advantage of using \benchname over raw performance counters (PCM) is that PCM cannot measure contention beyond the saturation point. For instance, the measured traffic saturates at the link bandwidth (e.g.,~85 GB/s on our testbed), while the contention will still increase due to queueing. With \benchname, we can distinguish between saturated and contended links. Our validation results in Section~\ref{sec:interference} confirm that using \benchname for quantifying interference can reach higher precision than raw performance counter measurement.

We also define a metric called \textit{interference coefficient} (denoted as $IC$) by running \benchname with one thread with 1 flop/element for a set number of iterations and measuring the relative runtime $T$ of \benchname. The interference coefficient is calculated as $\text{IC}=\frac{T}{T_\text{idle system}}$. The interference coefficient quantifies the interference induced by an application on the system.

\subsection{Emulation Platform}
We configure an emulation platform for the memory pool in Figure~\ref{fig:arch}. The methodology uses the socket interconnect in a dual-socket system to emulate a coherent disaggregated memory system. Similar methods are used in~\cite{maruf2023tpp,pond23,bsc_contention}. The default first-touch page allocation policy places allocations onto the local NUMA node until full, before spilling to other NUMA nodes (the remote memory).

The emulation platform uses an Intel Xeon testbed with two sockets and one NUMA node per socket. As illustrated in Figure~\ref{fig:profiler}, one socket represents a compute node, and the memory on the other socket represents a memory pool. The cores on the second socket are not used. The UPI interconnect between them represents the remote link. The intra-socket bandwidth is 73~GB/s and latency is 111~ns, while the inter-socket bandwidth is 34~GB/s and latency is 202~ns. The Linux kernel version is 5.14. For consistent results, we disable NUMA balancing and transparent huge pages (THP).

%For emulating multiple hosts, we used an eight-NUMA nodes AMD testbed with two sockets and six cores per NUMA node. The intra-socket bandwidth is 18~GB/s and latency is 185~ns, while the inter-socket bandwidth is 6~GB/s and latency is 524~ns. This platform can emulate up to four hosts (on socket 1) and up to four memory pools (on socket 2) connected by a common fabric.  Workloads on both systems use the same setup.%, only 8 cores are used on the Intel testbed.

%To characterize the testbeds, we measured the sustained memory bandwidth using STREAM Triad~\cite{} and the memory latency using a pointer-chasing benchmark~\cite{}.

%\todo{update with 6 threads. or remove AMD?}

%To emulate a host with a certain size of local memory, we fill the rest of the node with locked memory pages. Representing different potential system designs, we execute our experiments with 25\%, 50\% and 75\% of each workload's memory footprint backed by a memory pool.

\begin{figure}
    \centering
    \includegraphics[width=\linewidth]{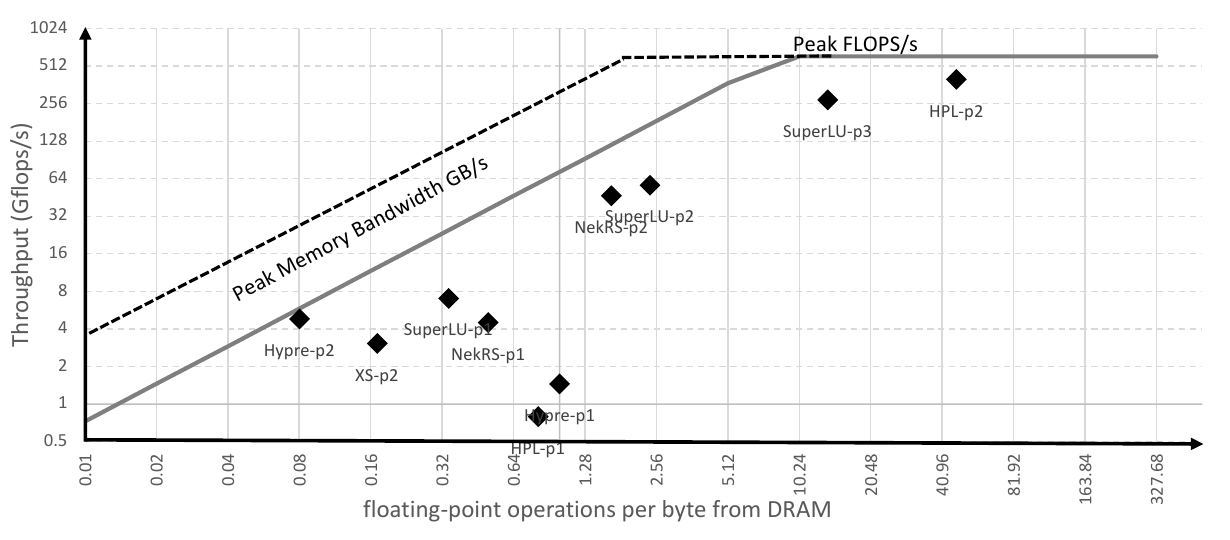}
    \caption{A roofline model built on our test platform. Dashed lines indicates extension by adding additional memory tiers.}
    \label{fig:roofline}
\end{figure}

\subsection{Analytical Models}\label{sec:roofline}
We use the standard roofline model and an extended memory roofline model to identify performance bottlenecks and hardware limits. The roofline model~\cite{williams_roofline_2009} was proposed to model the attainable peak performance~$P$ of a program with arithmetic intensity~$I$, defined as the number of floating-point operations per byte transferred from main memory. A computing platform is described by its peak computing power~$F$ in flop/s and peak memory bandwidth~$B$ in B/s, where $P = \min (F, B\cdot I)$. The solid line in Figure~\ref{fig:roofline} presents a standard roofline model derived from the architecture peak metrics (e.g., the peak flops from clock speed and AVX-512 vectors) and STREAM benchmark results. A roofline model can be further extended to model other limiting factors. For example, the slope of memory bandwidth can be adapted to include the impact of memory interference. The dashed line in Figure~\ref{fig:roofline} indicates an increase in total memory bandwidth if an additional memory tier is added to the baseline system.

For fine-grained measurement, we set our profiler to quantify the throughput in flop/s, and memory access in bytes every second. Figure~\ref{fig:roofline} reports the measured arithmetic intensity and obtained throughput for each phase in the evaluated applications in Table~\ref{tab:workloads}. The visualization of the measurement on the roofline model shows good coverage in arithmetic intensity and throughput. Therefore, we confirm that the evaluated workloads represent HPC workloads in the memory-bound to compute-bound spectrum. Furthermore, fine-grained measurement allows us to separate distinctive phases with different arithmetic intensities. As reported in Figure~\ref{fig:roofline}, each application typically consists of at least two phases, where the first phase (denoted as p1) represents the initialization phase. This experiment only uses the node-local memory and no other co-running applications. %Note that although most previous works only include the dominant phases for analysis. However, moving towards multi-tier memory, some access patterns in other phases may have an increased impact on the overall performance, and hence, we include all phases in this study.

%Figure~\ref{fig:roofline}, the framework measures the operational intensity for each parallel kernel and captures the list of kernels whose operation intensity is lower than the shaded region on the target platform. This step is a one-time effort performed offline and only requires host nodes and local memory. 

%Runtime monitoring is required to determine $B_{local}$ and $B_{remote}$ as they depend on data placement, access pattern, and cache policy.

\section{Workload Characterization}
In this section, we focus on the fundamental memory requirements of an application. We are interested in identifying and abstracting the properties that persist even when the application is executed on systems with different memory configurations. Table~\ref{tab:workloads} summarizes the list of evaluated applications and input problems used in this paper. Unless otherwise noted, experiments use the first listed input problem of each application.
\begin{table}[bt]
\begin{adjustbox}{width=\linewidth,center}{
\centering
\begin{tabular}{|l|l|c|c|c|c|c}
 \hline
 \textbf{Application} &\textbf{Description}  &\textbf{Parallelization} &\textbf{Input Problems}\\\hline\hline

\multirow{3}{*}{HPL~\cite{dongarra2003linpack}} 
    &\multirow{3}{*}{\shortstack[l]{High Performance LINPACK benchmark,\\dense LU factorization with partial pivoting.}}
    &\multirow{3}{*}{MPI+OpenMP} &N=20000\\ 
                                 & & &N=28280\\
                                 & & &N=40000\\\hline
\multirow{3}{*}{Hypre~\cite{hypre2}} 
    &\multirow{3}{*}{\shortstack[l]{Library of high-performance linear solvers.\\We use the structured interface.}}
    &\multirow{3}{*}{MPI+OpenMP} &\textsc{ex4} 10 times, n=6300, ranks=1\\ 
                                 & & &\textsc{ex4} 10 times, n=6300, ranks=2\\
                                 & & &\textsc{ex4} 10 times, n=6300, ranks=4\\\hline
\multirow{3}{*}{NekRS~\cite{fischer2021nekrs}} 
    &\multirow{3}{*}{\shortstack[l]{Computational fluid dynamics based on the\\ spectral element method.}}
    &\multirow{3}{*}{MPI} &turbPipePeriodic, p=5, dt=1e-2\\ 
                                 & & &turbPipePeriodic, p=7, dt=6e-3\\
                                 & & &turbPipePeriodic, p=9, dt=1e-3\\\hline

\multirow{3}{*}{BFS~\cite{shun2013ligra}} 
    &\multirow{3}{*}{\shortstack[l]{Graph processing benchmark of the breath-\\first search algorithm in the Ligra framework.}}
    &\multirow{3}{*}{OpenMP} &symmetric rMat, N=$2^{24}$, M=$2^{28.24}$\\ 
                                 & & &symmetric rMat, N=$2^{25}$, M=$2^{29.25}$\\
                                 & & &symmetric rMat, N=$2^{26}$, M=$2^{30.25}$\\\hline

\multirow{3}{*}{SuperLU~\cite{Li2005}} 
    &\multirow{3}{*}{Sparse LU factorization.}
    &\multirow{3}{*}{MPI+OpenMP} &SiO~\cite{Davis2011} (nnz=1.3M)\\ 
                                 & & &H2O~\cite{Davis2011} (nnz=2.2M)\\
                                 & & &Si34H36~\cite{Davis2011} (nnz=5.2M)\\\hline

\multirow{3}{*}{XSBench~\cite{tramm2014xsbench}} 
    &\multirow{3}{*}{\shortstack[l]{Monte Carlo neutron transport proxy\\application.}}
    &\multirow{3}{*}{MPI+OpenMP} &\textsc{large}, 2M particles, 11303 gridpoints\\ 
                                 & & &\textsc{large}, 2M particles, 22606 gridpoints\\
                                 & & &\textsc{large}, 2M particles, 45212 gridpoints\\\hline

%GUPS 		&scientific 	&\todo{xxx} 	& &MPI\\\hline
%stream		&scientific 	&\todo{xxx} 	& &MPI\\\hline
\end{tabular}
}\end{adjustbox}
\caption{Evaluated workloads with three input problems of approximately 1:2:4 memory usage ratio.}
\label{tab:workloads}
\end{table}

\begin{comment}
\begin{table}[bt]
\begin{adjustbox}{width=\linewidth,center}{
\centering
\begin{tabular}{|c|c|c|c|c|c|c|c}
 \hline
 \textbf{Application} &\textbf{Domain} &\textbf{Input Problem} &\textbf{Problem Scaling}  &\textbf{Parallelization} \\\hline\hline
HPL 	&scientific  	&$N=20000$   &$N$ &MPI+OpenMP\\\hline
Hypre  		&scientific 	&\textsc{ex4}, $n=6300$ &additional ranks &MPI+OpenMP\\\hline
NekRS 		&computational fluid dynamics  &turbPipePeriodic, $p=5$ &$p=7,p=9$ &MPI\\\hline
BFS 	&graph processing  			&symmetric rmat, $N=10M$  &$N$  &OpenMP\\\hline
SuperLU		&scientific 	&sparse matrix nnz=1.3M (SiO) 	&H2O, Si34H36 &MPI+OpenMP\\\hline
XSBench		&scientific 	&\textsc{large}, 2M particles 	&num. particles &(MPI)\\\hline
GUPS 		&scientific 	&\todo{xxx} 	& &MPI\\\hline
stream		&scientific 	&\todo{xxx} 	& &MPI\\\hline
\end{tabular}
}\end{adjustbox}
\caption{Workload and their characterization}
\label{tab:testbed}
\end{table}
\end{comment}

\subsection{Memory Capacity and Bandwidth Scaling}

Capacity and bandwidth are the two primary memory-related considerations when deploying an HPC application. In a typical decision flow, a user needs to estimate the total memory footprint of the job and peak memory usage per node, then compare them with memory capacity per compute node to determine the minimum number of nodes required. When memory bandwidth is a limiting factor, a user may decide to increase the number of nodes further for higher aggregate memory bandwidth. This may be guided by the roofline model. Other dimensions of this decision include increased communication and core-hour cost with more nodes.% However, over-decomposing into too many nodes could leave bandwidth and capacity underutilized with communication cost increased.

\begin{figure}
    \centering
    \begin{subfigure}[b]{0.325\linewidth}
        \centering
        \includegraphics[width=\textwidth]{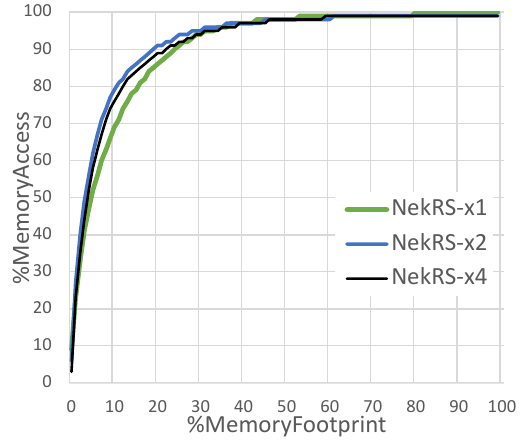}
        \caption{NekRS}\label{fig:scaling_nek}
    \end{subfigure}
    \begin{subfigure}[b]{0.325\linewidth}
        \centering
        \includegraphics[width=\textwidth]{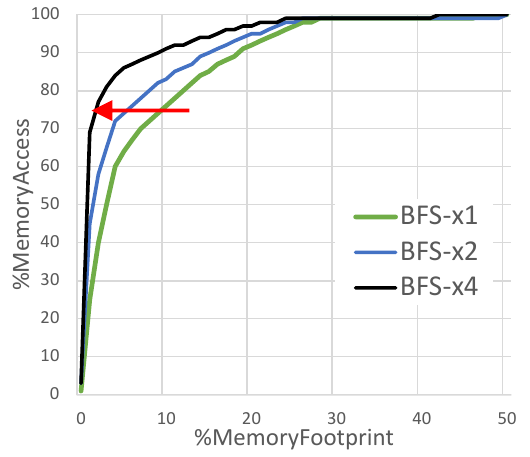}
        \caption{BFS}\label{fig:scaling_bfs}
    \end{subfigure}
    \begin{subfigure}[b]{0.325\linewidth}
        \centering
        \includegraphics[width=\textwidth]{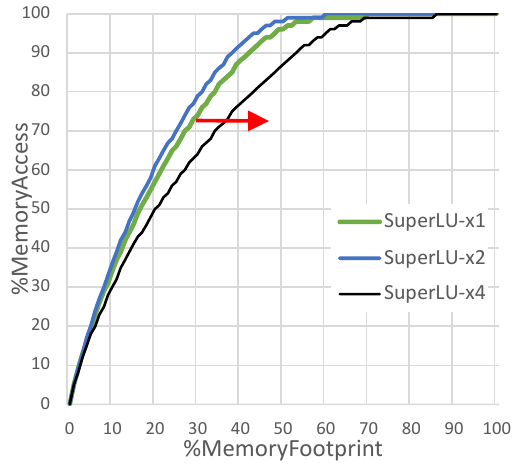}
        \caption{SuperLU}\label{fig:scaling_lu}
    \end{subfigure} 
    \begin{subfigure}[b]{0.325\linewidth}
        \centering
        \includegraphics[width=\textwidth]{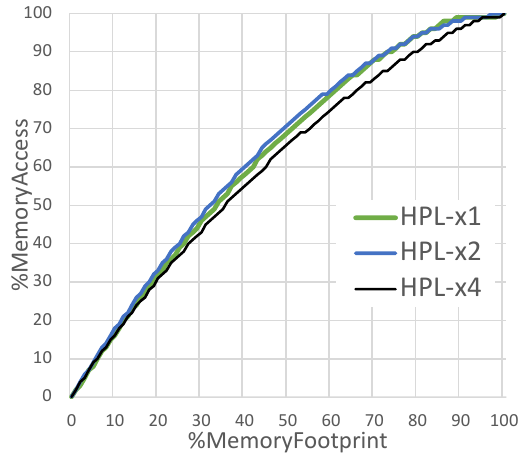}
        \caption{HPL}\label{fig:scaling_hpl}
    \end{subfigure}
    \begin{subfigure}[b]{0.325\linewidth}
        \centering
        \includegraphics[width=\textwidth]{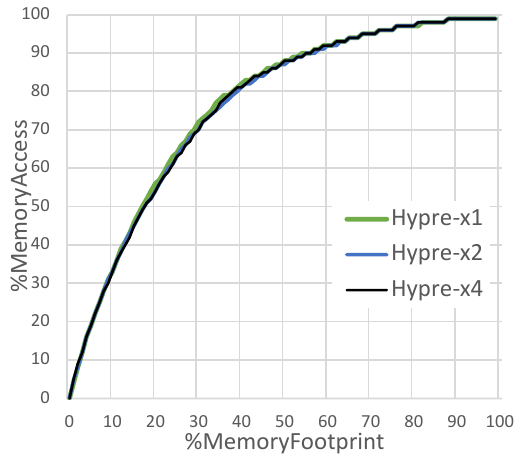}
        \caption{Hypre}\label{fig:scaling_hypre}
    \end{subfigure}
    \begin{subfigure}[b]{0.325\linewidth}
        \centering
        \includegraphics[width=\textwidth]{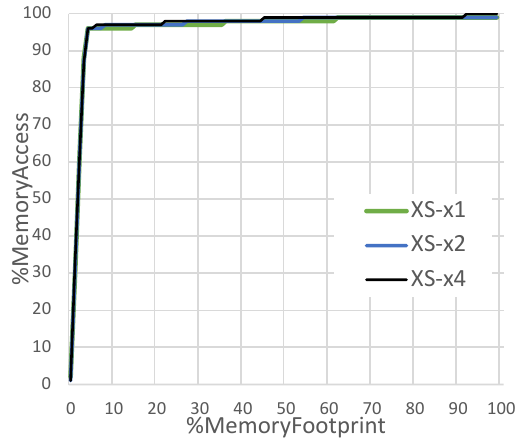}
        \caption{XSBench}\label{fig:scaling_xs}
    \end{subfigure}     
    \caption{The cumulative distribution function of memory accesses and memory footprint in six applications at three scaled input problems.}
    \label{fig:scaling}
\end{figure}

We propose a memory bandwidth-capacity scaling curve, as demonstrated in Figure~\ref{fig:scaling}, to help users quantify the relationship between capacity and bandwidth usage. The curve is built using the memory access sampling in our profiler. After measuring and aggregating the number of memory accesses by page, we sort pages into descending order of accesses. Then, we build the cumulative distribution of accesses to compare with the percentage of memory footprint. Each application is tested with three input problems of approximately doubling size.% For applications that use non-synthetic datasets (e.g., SuperLU), we choose input datasets with approximately doubling sizes. %Scientific applications tend to have larger objects~\cite{ji_understanding_2017}, and the major memory objects tend to grow as the input problems scale. 

The bandwidth-capacity scaling curve reveals that HPL and Hypre exhibit relatively uniform memory access across the memory footprint. Such characteristic is consistent with traditional numerical codes, where nearly all main memory objects are accessed for computation. In contrast, BFS and XSBench have only a small portion of the memory footprint being actively accessed during the execution. We check the source code and find that BFS allocates large graph structures, while only adjacency data will be accessed during execution. Similarly, XSBench allocates large grid structures while only sampled points will be looked up. 

A more interesting finding is that all applications but SuperLU exhibit a consistent bandwidth-capacity scaling pattern across increased input problems. Four applications, NekRS, HPL, Hypre, and XSBench, have overlapping scaling curves for $1\times$, $2\times$, and $4\times$ input problems, indicating their usage patterns in bandwidth and capacity preserve across input sizes. BFS has the scaling curve shifted towards the left side of the x-axis when the input problem increases, indicating the distribution of access is becoming more skewed, with fewer percentage of pages having more accesses.
%SuperLU is the only one that exhibits a flattening scaling curve. SuperLU used real sparse matrices as input, which have different sparsity structures.
The change in SuperLU indicates the distribution of hot pages moving from a skewed distribution towards a more uniform distribution. Overall, this scaling curve analysis is a concise visualization tool for HPC users to unify the two dimensions of memory requirements, i.e., bandwidth and capacity, and estimate requirements of larger-scale problems. 

%We quantify the ratio of main memory objects in NekRS in three input problems. As table~\ref{tab:ratio}, the sum of these major objects occupy more than 90\% of the total memory footprint. Also, as the input problem scales up, the major memory objects also scale up. As observed in scientific applications, major memory objects scale up with input problems. 

%We classify an application's memory footprint and bandwidth usage. A variable is considered major memory object if it attributes to most bandwidth usage or contribute to major memory footprint. A metric that unifies the two criteria is called \textit{access intensity}, often quantified as $\frac{Bytes_{accessed}}{Bytes_{size}}$. Intuitively, a higher access intensity corresponds to higher performance impact.

%\subsection{Suitability of Prefetching}
\subsection{Hardware Prefetching}

We consider the suitability of prefetching as a property determined by an application's algorithm and access pattern. Therefore, even if an application is executed on different systems, the suitability of prefetching is preserved. %, and the application could still benefit from hardware prefetching.
With disaggregated memory, prefetching may become even more important to hide the higher access latency. A benefit of non-faulting implementations, such as CXL, is that hardware prefetching can remain fully operational for remote memory.

To quantify the suitability of prefetching, we propose using two previously proposed metrics, namely \textit{Accuracy} and \textit{Coverage}~\cite{mahar2023workload}. \textit{Accuracy} measures the ratio of prefetched cachelines that have been actually accessed by the program. \textit{Coverage} measures the ratio of cacheline accesses that were prefetched before on-demand access. Our analysis in this work is based on L2 cache, where the core hardware prefetcher is located and LLC is an exclusive L3.

On our testbed, we set our profiler to capture measurements from four hardware counters: PF\_L2\_DATA\_RD, PF\_L2\_RFO, L2\_LINES\_IN, and USELESS\_HWPF. We calculate the two metrics as follows.
\begin{comment}
\begin{equation}
  Accuracy = \frac{pf\_l2\_data\_rd + pf\_l2\_rfo - useless\_hwpf}{pf\_l2\_data\_rd + pf\_l2\_rfo}
\end{equation}
\begin{equation}
  Coverage = \frac{pf\_l2\_data\_rd + pf\_l2\_rfo - useless\_hwpf}{l2\_lines\_in - useless\_hwpf}  
\end{equation}
\end{comment}
\begin{equation}
 \small\text{Accuracy} = \frac{\text{PF\_L2\_DATA\_RD} + \text{PF\_L2\_RFO} - \text{USELESS\_HWPF}}{\text{PF\_L2\_DATA\_RD} + \text{PF\_L2\_RFO}}\normalsize
\end{equation}
\begin{equation}
  \small\text{Coverage} = \frac{\text{PF\_L2\_DATA\_RD} + \text{PF\_L2\_RFO} - \text{USELESS\_HWPF}}{\text{L2\_LINES\_IN} - \text{USELESS\_HWPF}}\normalsize
\end{equation}

Additionally, we execute the workload with prefetching disabled to determine the performance gain prefetching contributes. Hardware prefetching is disabled by configuring a model-specific register in the processor core (the two least significant bits of MSR 0x1a4).

\begin{figure}
\centering
\includegraphics[width=\linewidth]{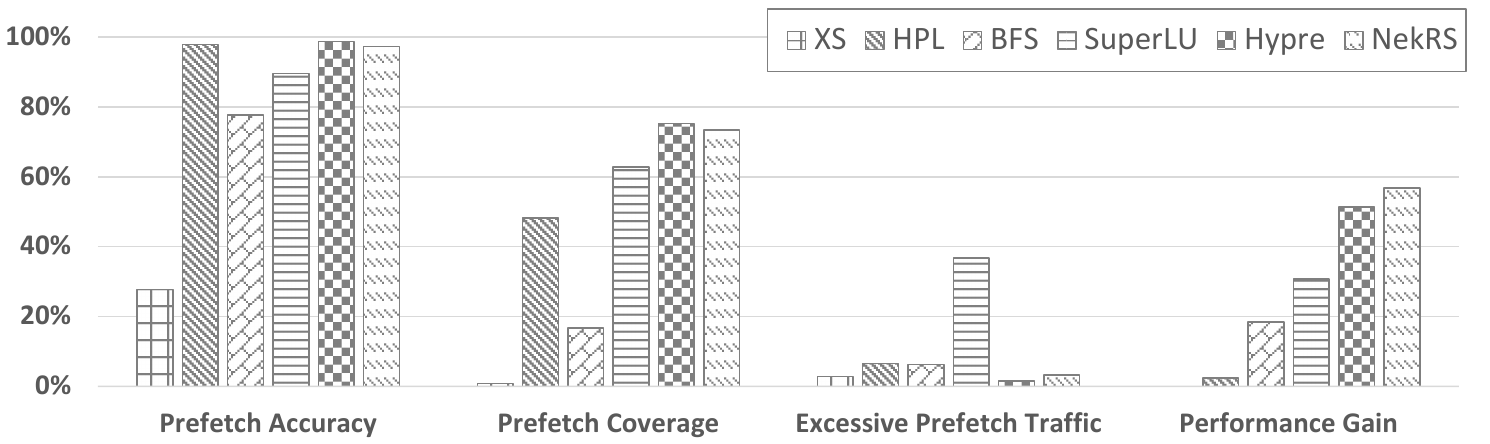}
\caption{Accuracy, coverage, excessive memory traffic, and performance gain from prefetching in all tested applications}
\label{fig:prefetch2}
\end{figure}

Figure~\ref{fig:prefetch2} reports the measured prefetching accuracy and coverage of each application. All except XSbench and BFS have more than 80\% prefetching accuracy. Hypre and NekRS have the highest prefetching coverage -- about 70\% of L2 cacheline accesses are prefetched instead of fetched on demand. Figure~\ref{fig:prefetch2} presents a timeline of the number of fetched cachelines with and without L2 prefetching in NekRS, HPL, and XSBench, respectively. 
\begin{figure}
    \centering
    \begin{subfigure}[b]{0.325\linewidth}
        \centering
        \includegraphics[width=\textwidth]{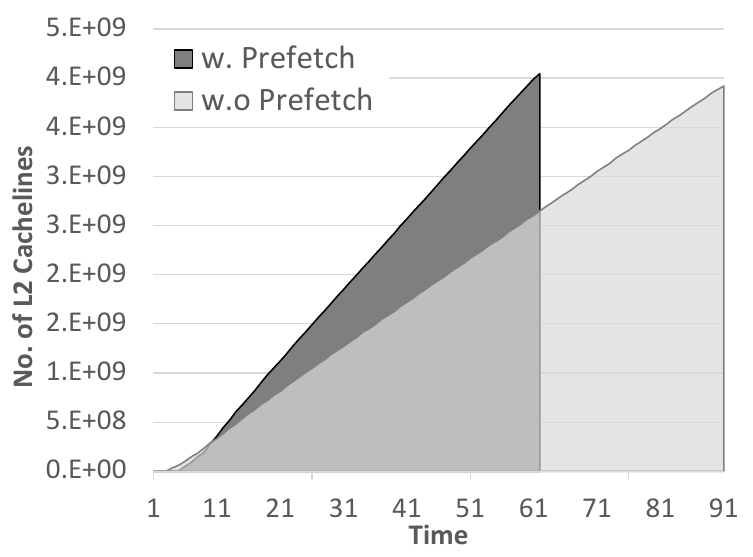}
        \caption{NekRS}\label{fig:pf_nek}
    \end{subfigure}
    \begin{subfigure}[b]{0.325\linewidth}
        \centering
        \includegraphics[width=\textwidth]{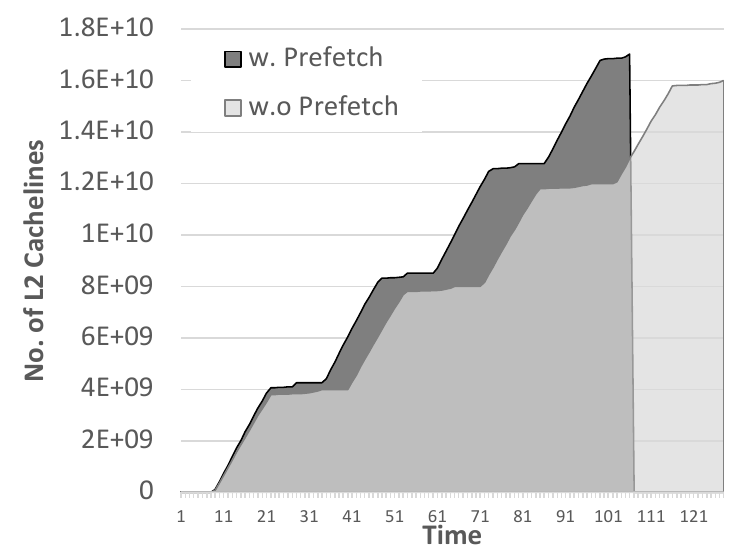}
        \caption{HPL}\label{fig:pf_hpl}
    \end{subfigure}
    \begin{subfigure}[b]{0.325\linewidth}
        \centering
        \includegraphics[width=\textwidth]{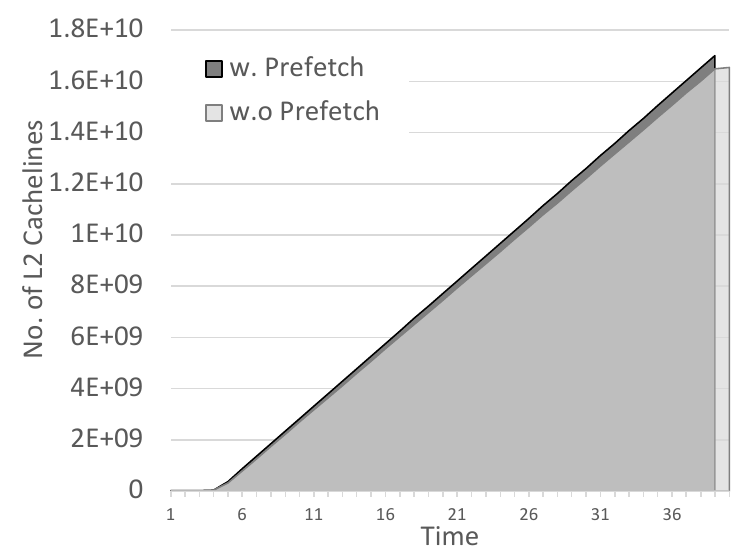}
        \caption{XSBench}\label{fig:pf_xs}
    \end{subfigure}
    \caption{Measured memory traffic in number of cachelines (y-axis) and runtime (x-axis) in three applications with and without L2 prefetching enabled.}
    \label{fig:prefetch1}
\end{figure}    
The results show that prefetching may contribute to a substantial portion of memory bandwidth in the tested applications. For instance, Figure~\ref{fig:pf_nek} shows that the memory bandwidth consumption in NekRS is significantly higher when prefetching is enabled. Note that the total memory traffic in NekRS with prefetching is only 3\% higher than that without prefetching, but the performance gain is as high as 57\%, as reported in Figure~\ref{fig:prefetch2}. A similar observation of prefetching contributing substantial memory traffic is also identified in cloud workloads~\cite{mahar2023workload}. However, these workloads exhibit low prefetching accuracy and coverage, and thus, prefetching is considered harmful. In contrast, the high coverage and accuracy, and performance gain in our experiments indicate the suitability of prefetching for scientific workloads. 

All applications have low excessive memory traffic ($2\%$-$6\%$ as reported in Figure~\ref{fig:prefetch2}) due to prefetching, except SuperLU where the total memory traffic with prefetching enabled is 37\% higher than that with prefetching disabled. However, the performance gain is almost 31\% and thus prefetching is still useful from a performance perspective. Although XSBench has the lowest accuracy, it also has low excessive memory traffic ($3\%$) from prefetching, indicating the prefetching is automatically adapted to a low level when accuracy is low.

%Page-level prefetching for DM~\cite{hopp2023}.
%\todo{combine multiple workloads on 1 node to show if it confuses the prefetching in contrast with Google's paper}

\begin{tcolorbox}[boxsep=2pt,left=2pt,right=2pt,top=2pt,bottom=2pt]
  The memory bandwidth-capacity scaling curve unifies two main factors and aids users in projecting memory usage on different problems. 
  Prefetching may constitute major memory usage and is critical for the performance of HPC applications.% and may become critical for bridging remote memory accesses. 
\end{tcolorbox}
\section{Multi-tier Memory}
%\section{Transforming towards Multi-tier Memory}
\begin{figure*}
    \centering
    \begin{subfigure}[b]{0.325\textwidth}
        \centering
        \includegraphics[width=\textwidth]{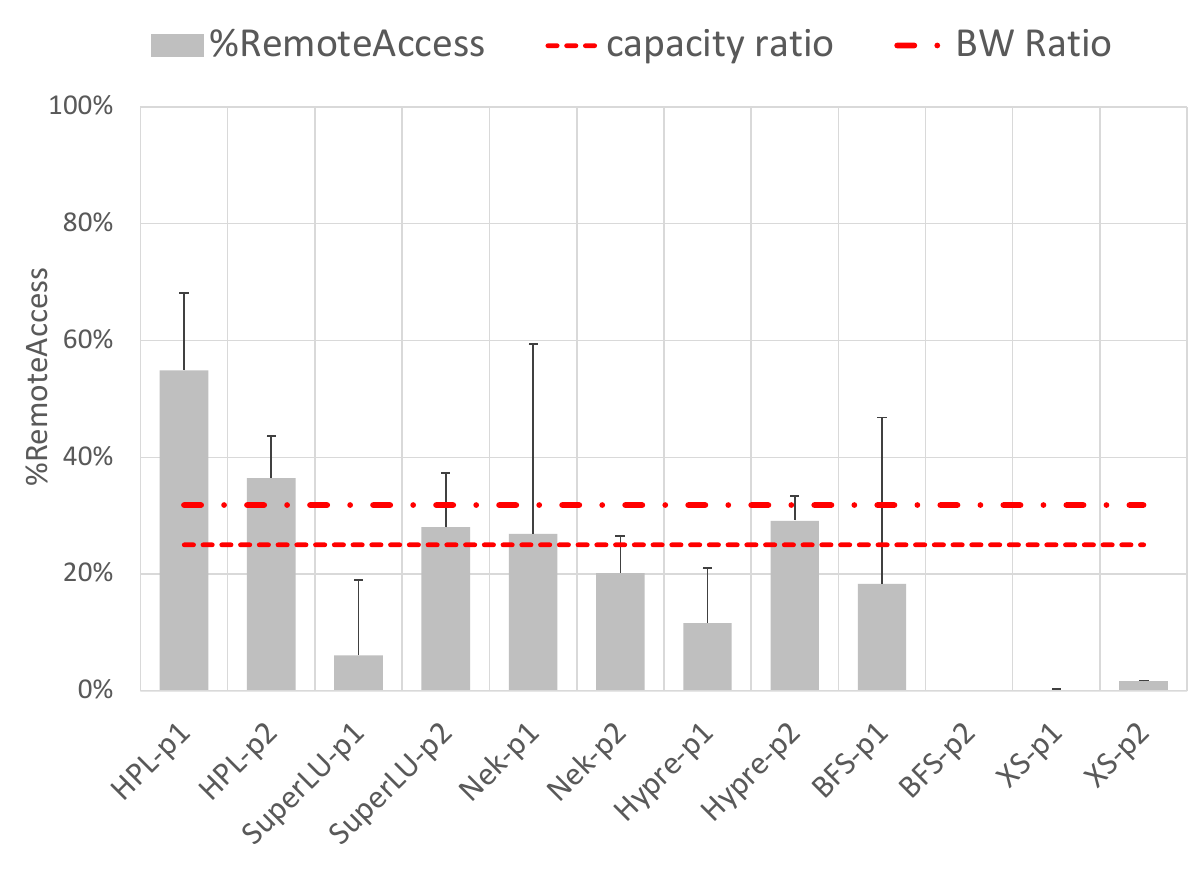}
        \caption{75\%-25\% capacity ratio}\label{fig:traffic25}
    \end{subfigure}
    \begin{subfigure}[b]{0.32\textwidth}
        \centering
        \includegraphics[width=\textwidth]{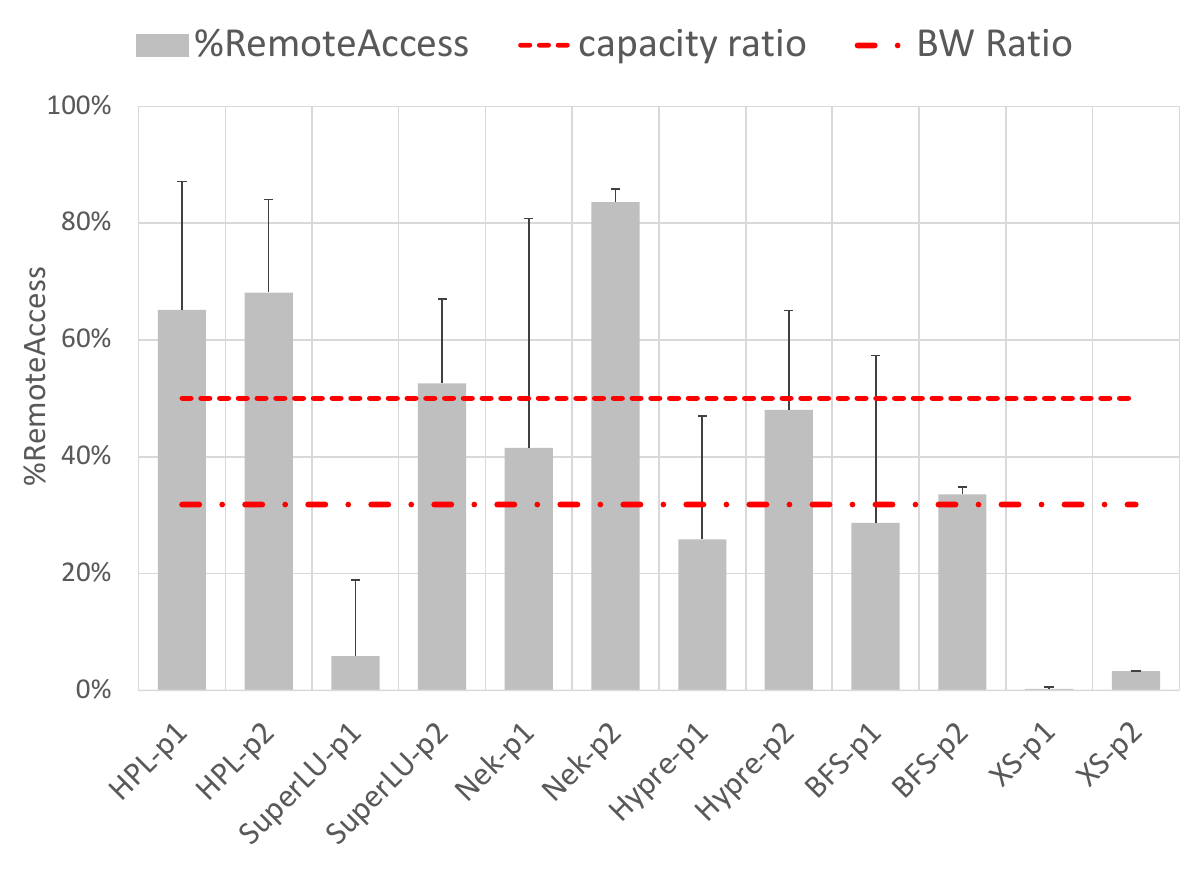}
        \caption{50\%-50\% capacity ratio}\label{fig:traffic50}
    \end{subfigure}
    \begin{subfigure}[b]{0.32\textwidth}
        \centering
        \includegraphics[width=\textwidth]{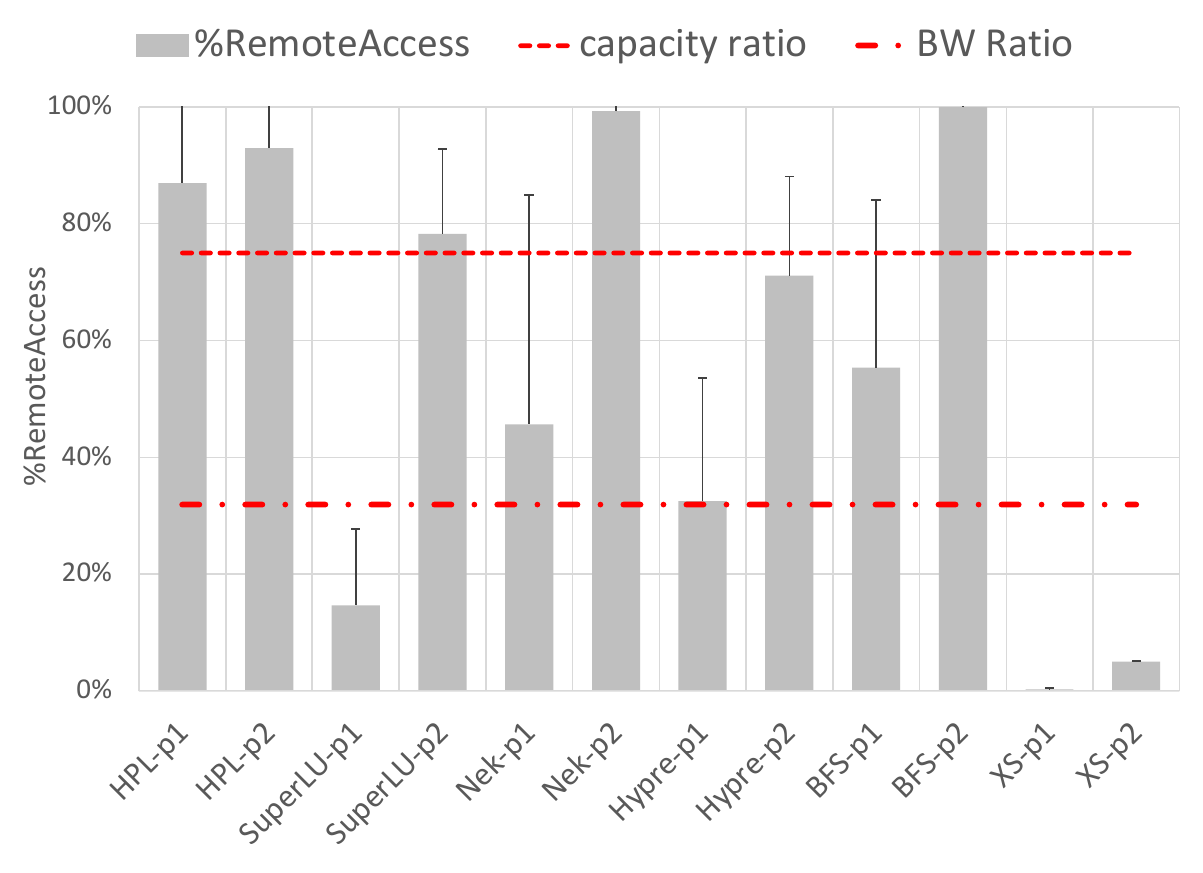}
        \caption{25\%-75\% capacity ratio}\label{fig:traffic75}
    \end{subfigure}
    \caption{The ratio of memory accesses to the second tier on three two-tier memory systems with capacity ratios ranging from 25\% to 75\%. Label X-pY denotes the Y-th phase of workload X.}
    \label{fig:traffic}
\end{figure*}

In this section, we cover the application impact when introducing an additional memory tier. A general multi-tier memory system has relatively faster but smaller top tiers and slower but larger bottom tiers. The exact memory technologies for implementing each tier may vary. For the experiments in this section, we emulate a two-tier memory system and run the profiler on selected applications to quantify memory access to each tier. 

To capture the impact of multi-tier memory on performance bottlenecks, a memory roofline model was proposed in previous work~\cite{lbl_dm} to model the memory access performance as a function of the local to remote memory access (L:R) ratio. The model guides tuning towards high L:R ratios to shift the limit of the memory access performance toward the peak bandwidth of the fast-tier memory. In fact, the peak memory performance can be further increased by leveraging all memory tiers concurrently instead of solely the fast tier. Therefore, we emphasize balanced local to remote accesses that match the bandwidth and capacity limit of each memory tier. 

\subsection{Tiered Memory Access}
%When the memory footprint of an application is larger than the memory capacity of the compute node (host), extra memory resources will be provisioned from a memory pool (memory server). Therefore, the memory bandwidth usage of a kernel is composed of two parts: from the memory local to the cores running the kernel and from the memory remote to the cores. Thus, the total bandwidth is the sum $B = B_{local} + B_{remote}$. 

Two reference points are critical for guiding optimization of applications on multi-tier memory systems. The first one is the ratio of the memory capacity of each tier. The second reference point is the ratio of memory bandwidth of each tier. First, using our profiler, we quantify the ratio of memory accesses to each memory tier. Then, we calculate $R_{Cap}^i$, $R_{BW}^i$, and $R_{access}^i$, respectively, to quantify the ratio of memory capacity, bandwidth, and access to a tier $i$ and compare them with the two reference points.

Figure~\ref{fig:traffic} reports the measured $R_{access}^i$ of each phase in the tested applications on three system configurations, where $R_{Cap}^i=25\%, 50\%, 75\%$. As a validation of our profiler, we also measured the arithmetic intensity of each phase identified in Section~\ref{sec:roofline} by using $AI=\frac{FLOPS}{Byte_{LM} + Byte_{RM}}$, where $Byte_{LM}$ denotes the number of bytes retrieved from local memory tier and $Byte_{RM}$ denotes the number of bytes retrieved from remote memory tier. The measured arithmetic intensities are consistent with those reported in the experiment on the single-tier system in Figure~\ref{fig:roofline} and are omitted due to the space limit.

Figure~\ref{fig:traffic} reports the percentage of accessed bytes from the remote memory tier among all memory accesses, i.e., $R_{access}^{remote}$. We also add two reference lines of $R_{BW}^{remote}$ and $R_{Cap}^{remote}$. $R_{BW}^{remote}$ is the turning point of memory access bottleneck, where a $R_{access}^{remote}$ value lower than $R_{BW}^{remote}$ indicates that the memory access bottleneck is bound by the bandwidth of the fast tier. A $R_{access}^{remote}$ value higher than $R_{BW}^{remote}$, however, indicates too many memory accesses to the slower tier so that the slower tier becomes the limit of memory access performance.  Thus, the $R_{BW}$ reference is an upper bound for tuning. The lower bound of tuning uses the reference of memory capacity ratios $R_{Cap}$. The ratio of memory accesses to each tier should at least match the ratio of their capacity. 

%We can identify different optimization priorities on a general multi-tier memory system by leveraging the two reference points and the measured memory access metrics from the profiler.
By leveraging the two reference points and the memory access metrics from our profiler, optimization opportunities on multi-tier memory can be identified and prioritized.
In Figure~\ref{fig:traffic25}, the two reference lines are close, and most applications (except HPL and XSBench) have their ratios of remote memory access close to the two optimization reference lines, indicating little optimization space, and users should not spend efforts in optimizing data placement. In contrast, Figure~\ref{fig:traffic75} presents plenty of optimization opportunities from the two reference lines. For instance, based on the capacity ratio, HPL, NekRS, and BFS all exhibit excessive accesses above the $R_{Cap}$ (denoted in the dashed red line). The second phase (p2) of almost all applications is substantially above the bandwidth ratio ($R_{BW}$) reference. Although this indicates a large space for optimization, it could also mean an ill-balanced design of the memory tiers. Note that although all phases are included, not all carry the same weight on the total execution time.

XSBench stands out among all applications in that the remote access ratio is low (below 6\%) in all configurations. This means that the additional bandwidth from remote memory is not utilized. However, considering that the prefetching coverage for XSBench is $<1\%$, the application is highly sensitive to increased memory latency. Therefore, reducing latency by minimizing remote memory exposure  is more important than increasing the aggregate memory bandwidth.

\subsection{Phase Changes in Memory Access}
HPC applications often consist of several phases. As illustrated in Figure~\ref{fig:roofline} and Figure~\ref{fig:traffic}, phases in an application can have very different profiles of arithmetic intensity and memory access patterns. This characteristic increases the complexity of optimizing applications on multi-tier memory systems. For instance, one scheme for a kernel's data placement may harm other kernels. This global optimization is a Knapsack problem which is NP-complete. 

Static solutions leverage offline profiling to modify allocation sites to place a memory variable directly on a suitable memory tier. Dynamic solutions resort to runtime detection of access patterns to migrate performance-critical memory pages into the fast tier. The first direction requires extensive porting efforts and is often not adaptable to new input problems or system architecture. The second direction is user transparent. However, performance variation and indeterministic performance are often unacceptable to HPC end users. Also, on high-end HPC hardware, phases in an application are often short in time, while runtime solutions need time to adapt.

%While existing kernel-based solutions on runtime migration of hot pages into the local memory, such solutions have two main limitations. First, the frequent phase changes cause frequent data movement between memory tiers. Second, relying on the kernel to pick up the changes in access patterns, e.g., hot pages in a phase, takes time and incurs extra monitoring overhead.

\begin{tcolorbox}[boxsep=2pt,left=2pt,right=2pt,top=2pt,bottom=2pt]
%  The upper bound of memory traffic distribution to each tier should match the capacity ratio between memory tiers.
Capacity ratio and bandwidth ratio provides two optimization references and the ratio of memory access to tiers should match them.
The dominant phase with unmatched memory access distribution should be the priority of optimization. 
\end{tcolorbox}

\section{Interference on Memory Pooling}\label{sec:interference}
\begin{figure*}
    \centering
    \begin{subfigure}[b]{0.325\textwidth}
        \centering
        \includegraphics[width=\textwidth]{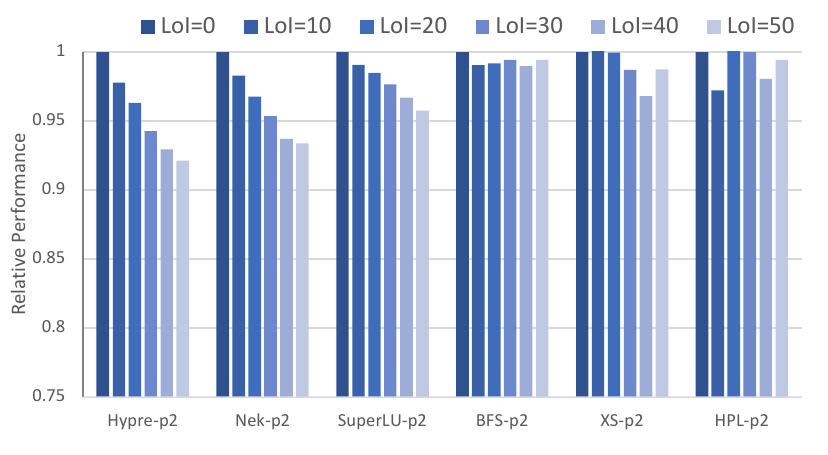}
        \caption{25\%-75\% capacity ratio}\label{fig:sensitivity25}
    \end{subfigure}
    \begin{subfigure}[b]{0.325\textwidth}
        \centering
        \includegraphics[width=\textwidth]{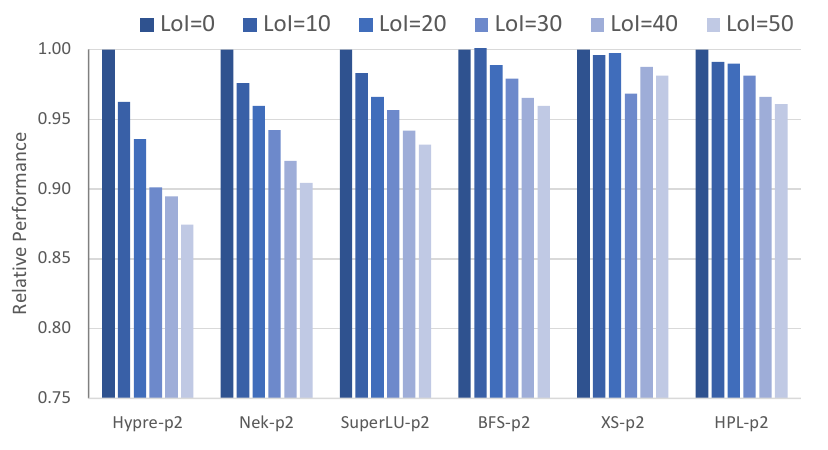}
        \caption{50\%-50\% capacity ratio}\label{fig:sensitivity50}
    \end{subfigure}
    \begin{subfigure}[b]{0.325\textwidth}
        \centering
        \includegraphics[width=\textwidth]{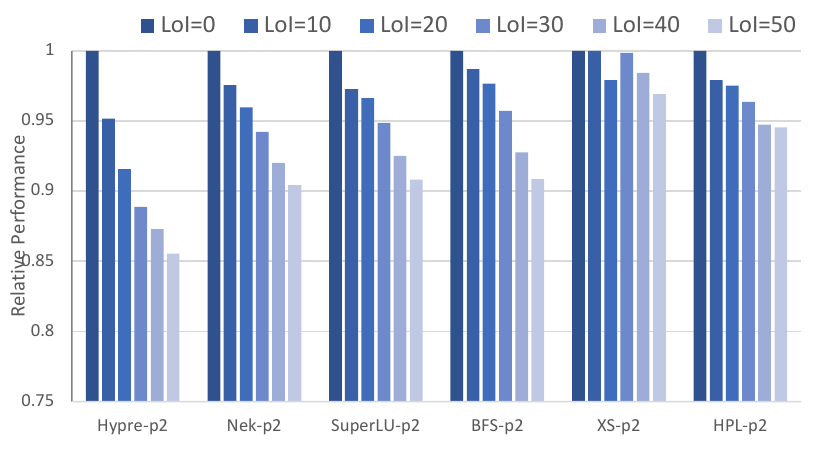}
        \caption{25\%-75\% capacity ratio}\label{fig:sensitivity75}
    \end{subfigure}
    \caption{Quantifying application's sensitivity to memory interference on three disaggregated memory systems based on memory pool. Y-axis indicates the relative performance w.r.t LoI=0.}
    \label{fig:sensitivity}
\end{figure*}
\begin{figure}[bt]
    \centering
    \begin{subfigure}[b]{0.3\linewidth}
        \centering
        \includegraphics[height=2.5cm,width=\textwidth]{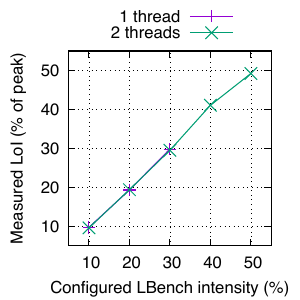}
        %\caption{scaling}
    \end{subfigure}
    \begin{subfigure}[b]{0.35\linewidth}
        \centering
        \includegraphics[height=2.5cm,width=\textwidth]{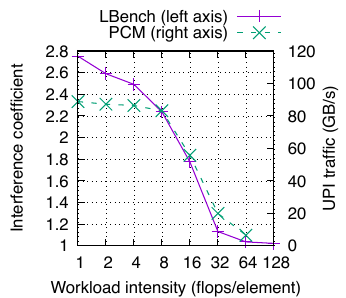}
        %\caption{accuracy}
    \end{subfigure}
    \begin{subfigure}[b]{0.325\linewidth}
        \centering
        \includegraphics[height=2.5cm,width=\textwidth]{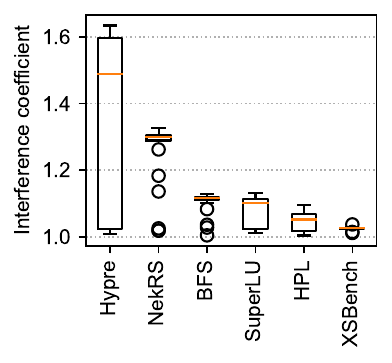}
        %\caption{Interference}
    \end{subfigure}  
    \caption{The scaling of interference using our benchmark (left), the comparison of measurement from our benchmark and raw performance hardware counter (middle), the interference caused by different applications (right).}
    \label{fig:interference}
\end{figure}

In this section, we focus on memory pool-based disaggregated memory systems. As introduced in Section~\ref{sec:bg}, disaggregated memory is a specific form of multi-tier memory. In the example in Figure~\ref{fig:arch}, the bottom memory tier is implemented by a memory pool, and the top memory tier is implemented by each node's local memory. As multiple nodes may share the memory pool, a unique challenge is memory interference from co-running jobs on other nodes. We use \benchname to quantify the impact of memory interference the applications in Table~\ref{tab:workloads}. For each application, we quantify two aspects of memory interference -- its sensitivity to memory interference and the memory interference caused by the application. The first metric is important for HPC users to estimate their application's performance on future HPC systems equipped with the pool-based memory tier. The second metric is useful for schedulers to improve co-location decisions.
%we propose two metrics called $Intf_{sen}$ and $Intf_{ind}$. $Intf_{sen}$ quantifies an application's sensitivity to memory interference while $Intf_{ind}$ quantifies the interference induced by an given application.

The first step of our study is to validate the accuracy and precision of \benchname. Thus, we configure the benchmark intensity to sweep from 10 to 50 and measure the percentage of link traffic compared to the peak. As reported in the left panel of Figure~\ref{fig:interference}, the measured level of intensity (y-axis) is linearly proportional to the configured intensity in \benchname. Thus, we confirm \benchname can be used to generate the required level of interference. Next, we compare the precision of \benchname with the low-level hardware counters (denoted as $PCM$). We set the background workload to sweep arithmetic intensity from 1 to 128 flops per element. Then we measure the resulting interference coefficient ($IC$) as the relative runtime of \benchname and report it in the middle panel of Figure~\ref{fig:interference}. Additionally, we measure the raw link traffic from PCM. The results show that at
%low workload intensity 
high bandwidth usage
(i.e., below $8$ flops/element), \benchname can still quantify 
%interference coefficients at $0.2$ precision
increased contention
while the hardware counter measurement saturates around 85~GB/s. The improved precision is because contention can continue increasing due to queueing effects while the measured traffic saturates at the link bandwidth. Hence, we show that \benchname can provide accurate measurements and injection of a configurable level of intensity ($LoI$).

For the study in this section, we configure \benchname to run with two threads for interference generation as it provides up to 50\% intensity. The threads are running on the local socket to inject congestion on the link to the remote memory. We use only two threads for injection to mitigate the impact of shared L3 caches with the co-running workloads. The performance impact at 50\% intensity was measured to be less than 1\% for all workloads. 

%\todo{We validate the accuracy of the link benchmark by measuring the relative performance of our workloads running \textit{with host memory only}, while injecting link interference using the benchmark. Measurement of interference by timing the benchmark compared with UPI traffic measured by PCM. In the background, a workload with varying arithmetic intensity is executed.}

\subsection{Sensitivity to Memory Interference}\label{sec:intf_sens}

We quantify an application's sensitivity to memory interference on memory pooling using \benchname. We configure \benchname to generate interference of increased intensity levels, i.e., $LoI=0, 10, ...$. Then, for each application, we use its performance at $LoI=0$ as the baseline and its relative performance at higher $LoI$ as the measurement of its sensitivity to interference. A similar approach is also used in previous work~\cite{wangcanvas2022}. 

Figure~\ref{fig:sensitivity} reports the measured sensitivity of each application to interference on the memory pool on three system configurations, i.e., the ratios between local memory capacity and remote memory pool ranges from $25\%$, $50\%$, and $75\%$. In general, all applications show reduced performance at an increased level of memory interference. Hypre and NekRS are among the most sensitive applications to memory interference. On the 50-50\% tiers setup, Hypre and NekRS show 15\% and 13\% performance loss at $LoI=50$. In Figure~\ref{fig:traffic}, the memory access to the remote tier (i.e., memory pooling in this case) is not the highest compared to other applications. However, due to their low arithmetic intensity, as quantified in Figure~\ref{fig:roofline}, they show higher sensitivity than other applications. In contrast, results in Figure~\ref{fig:traffic} show that HPL has high accesses to the memory pool, while having a low sensitivity to memory interference. On the 50-50\% tiers setup, it shows less than $5\%$ performance loss even at the highest level of interference.

An application's sensitivity to memory interference on memory pooling is caused by its remote memory access and is inversely influenced by its arithmetic intensity. A quantification of application sensitivity to memory interference on a target disaggregated memory system is necessary for application users to determine its deployment configurations on the system. For applications with low sensitivity to memory interference, users can configure the job deployment to leverage more capacity from the memory pool and reduce the number of compute nodes needed to support the job. For highly sensitive applications, the users can choose to deploy a job with more compute nodes to reduce the remote memory access or even completely avoid remote memory to fulfill their performance requirements. Our quantification method provides a tool for HPC end users to make informed decisions and tradeoffs, improving user confidence in the new system architecture. 

%{Performance Variations} \todo{showing boxer whisker plot of (1) a set of 4 applications co-running with each other (2) or with injected iBench in the background }
%\textbf{Correlation with AI and Remote Access, Pearson}

\subsection{Interference Coefficient}
We propose a second metric, the \textit{Interference Coefficient}, to measure the interference caused by an application on the memory pool. As an application accesses the remote memory tier, it also injects memory traffic onto the shared pool, causing interference on the system and other jobs. The main difference from the first metric, i.e., sensitivity, is that this metric is solely related to the remote memory access but is not directly influenced by the arithmetic intensity of the application. As the sensitivity of an application is important for HPC end users, this metric is important for system-level coordination. For example, the interference coefficient could be provided as a hint in job descriptions to enable schedulers with interference awareness to improve co-location decisions.

We quantify the interference coefficient of applications on a target system by co-running the application with \benchname and calculating the relative slowdown compared to an idle system. The right panel of Figure~\ref{fig:interference} reports the measured interference coefficient for the applications on a 50\% memory pooling setup. The results show NekRS and Hypre can introduce the most memory interference to other co-running jobs while HPL and XSBench introduce the lowest interference. The spread in Figure~\ref{fig:interference} highlights the variance in the interference coefficient caused by a workload. For instance, in Hypre, the compute phase causes high interference coefficient, while the initialization phase causes a low interference coefficient. 

%\textbf{(MPI) Induced Interference} MPI. It's an experiment checking the hypothesis that MPI synchronization would lead to higher performance impact. The setup is two Nek jobs, each with 2 nodes (16 ranks). They have 75\% pool residency. Either each job is isolated to 1 pool, or they are mixed so that both use both pools. The results show that mixing is 4\% faster, indicating that the hypothesis is actually true. I think it is an interesting result and has implications for scheduling. But I’m not sure if we can call it a “use case”?

\begin{tcolorbox}[boxsep=2pt,left=2pt,right=2pt,top=2pt,bottom=2pt]
  An application's sensitivity to memory interference is determined by the arithmetic intensity and the amount of accesses to memory pools. Users need to incorporate the sensitivity to memory interference in deployment decisions.
\end{tcolorbox}
\section{Use Cases}
In this section, we demonstrate use cases for leveraging the results of the three-level quantitative study to guide optimization at both the application level and the system level.

\begin{figure}
    \centering
    \includegraphics[width=\linewidth]{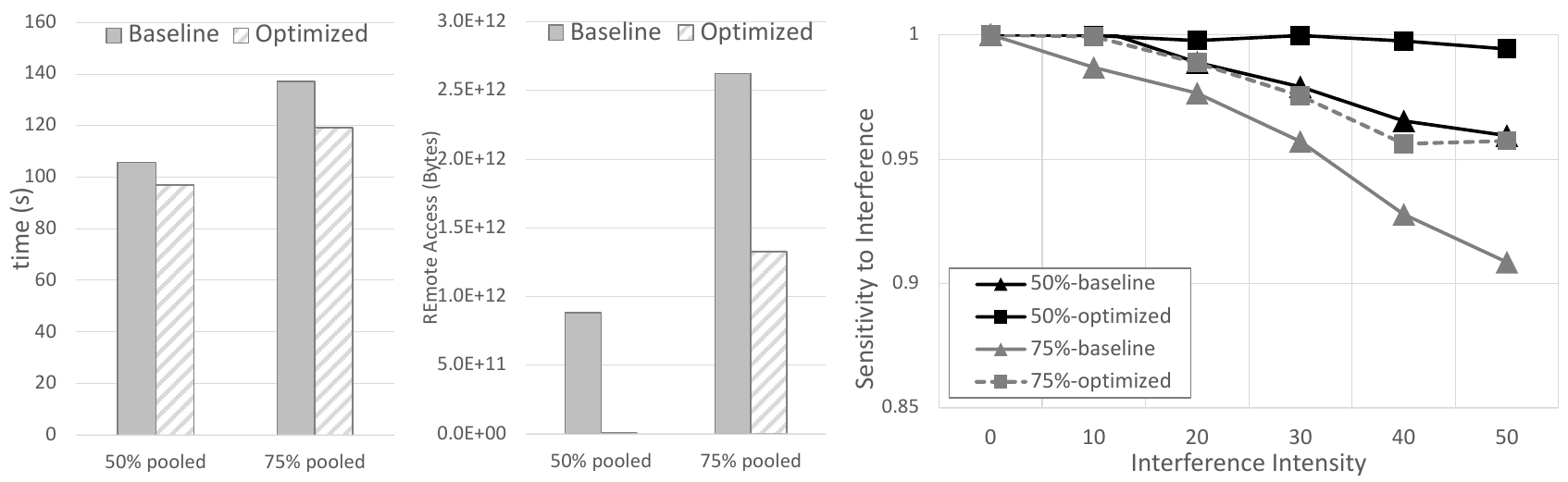}
    \caption{Optimization on data placement in BFS improves runtime, reduces remote memory accesses and sensitivity to memory interference.}
    \label{fig:bfsopt}
\end{figure}

\subsection{Optimizing Remote Memory Traffic}
In the first case study, we show that analyzing the memory access distribution on memory tiers can guide programmers to prioritize the optimization of data placement at the application level. %\begin{equation}\frac{Traffic_{remote}}{Traffic_local} > \frac{Mem_{pool}}{Mem_local}.\end{equation}
The results in the multi-tier analysis of BFS with 75\% remote capacity showed 99\% remote memory access -- the remote access ratio is much higher than the capacity ratio reference. The mismatch indicates that the most accessed data structures are in the remote memory. Coupled with information obtained from memory allocation sites in our profiler, we identified major memory objects and found that the \texttt{Parents} array is small but highly accessed. 

We consider three options to change its placement into the node-local memory. First, explicitly allocate in local memory, which is straightforward with libnuma. However, in the case of BFS, several large objects are allocated before \texttt{Parents}, leaving no space in local memory at the allocation site of \texttt{Parents}. Alternatively, we can explicitly allocate less accessed objects in remote memory to free up space.
%However, this may unnecessarily increase the remote residency ratio as the local memory may become underutilized. 
The second option is to migrate to local memory after initialization by exchanging pages between the local and remote memory using libnuma. However, this is not feasible to implement at the application level since it requires information on all other memory objects in local memory, such as size and access intensity. Thus, this option is often implemented at the system software level. The third option is application specific, by changing the order of allocations. By allocating and initializing objects in order of hotness, the hottest objects will be placed in local memory due to the first-touch policy. Note that this option is only suitable for applications where all memory objects have a similar lifespan (usually the whole program duration). With frequent dynamic allocations and deallocations, this option may cause the local memory to be under-utilized.

We choose to change the allocation order in BFS so that \texttt{Parents} is allocated and initialized first. With this simple change, the remote access ratio is reduced from 99\% to 80\% in the 75\% remote memory scenario, resulting in a 6\% performance improvement. Further re-applying the multi-tier memory traffic analysis shows that now many remote accesses come from dynamic heap allocations. In BFS, these are used for temporary structures, including the current frontier. Since they are dynamic in scope and size, they cannot be pre-allocated like the parents. We attempted to reserve space in the local memory by allocating a block of memory at the start and freeing it at the end of the initialization. However, this did not yield a significant benefit.

Instead, by inspecting the code, we identified a temporary object used during initialization but not afterward. The original code leaves the object unfreed due to a performance bug in the (de)-allocator. 
%The free call was commented out by the original developers to work around a performance bug in the allocator. 
Indeed, freeing the object degrades performance by 3\% when running in local memory only. However, with the memory pool as the secondary tier, freeing up the object reserves more local memory for dynamic allocations, offsetting the performance overhead of deallocation. As shown in Figure~\ref{fig:bfsopt}, this 1-line change further reduced the remote access ratio from 80\% to 50\%, resulting in a 13\% performance improvement over the baseline at 75\% memory pooling. At 50\% pooling, the optimized version almost eliminates remote memory access as shown in the middle panel of Figure~\ref{fig:bfsopt}. We re-evaluate the application's sensitivity to memory interference as in Section~\ref{sec:interference} using the optimized version for the cases of 50\% and 75\% pooling. We compare the interference sensitivity in the right panel of Figure~\ref{fig:bfsopt}. The results show that the optimized version has a significantly reduced interference sensitivity. %The peak working-set size was reduced by 2\%. To balance the degradation without remote memory and the improvement with remote memory, a heuristic could be used to free the memory only if insufficient free local memory is available.
In the general case, the interfaces provided by Linux are insufficient for this kind of data placement optimization. An interface for reserving a portion of local memory for dynamic allocation would empower developers to optimize their applications for disaggregated memory.

%\todo{describe the limitation of current kernel or allocator solutions}

%New idea: in Ligra init, check file size before reading, if >X\% of free local memory, allocate file data in remote directly

\subsection{Interference-Aware Job Scheduling}
\begin{figure}
    \centering
    \begin{subfigure}[b]{0.325\linewidth}
        \centering
        \includegraphics[width=\textwidth]{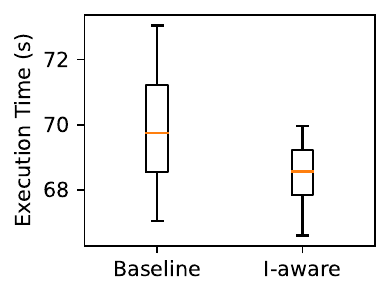}
        \caption{NekRS}\label{fig:corun_nek}
    \end{subfigure}
    \begin{subfigure}[b]{0.325\linewidth}
        \centering
        \includegraphics[width=\textwidth]{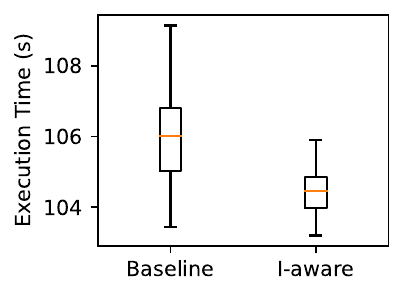}
        \caption{BFS}\label{fig:corun_bfs}
    \end{subfigure}
    \begin{subfigure}[b]{0.325\linewidth}
        \centering
        \includegraphics[width=\textwidth]{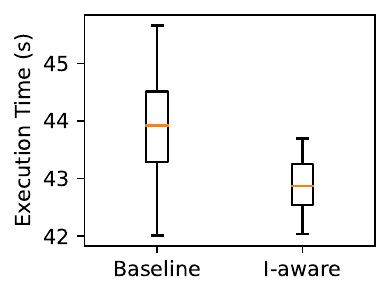}
        \caption{SuperLU}\label{fig:corun_superlu}
    \end{subfigure} 
    \begin{subfigure}[b]{0.325\linewidth}
        \centering
        \includegraphics[width=\textwidth]{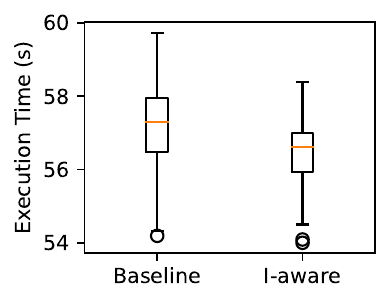}
        \caption{HPL}\label{fig:corun_hpl}
    \end{subfigure}
    \begin{subfigure}[b]{0.325\linewidth}
        \centering
        \includegraphics[width=\textwidth]{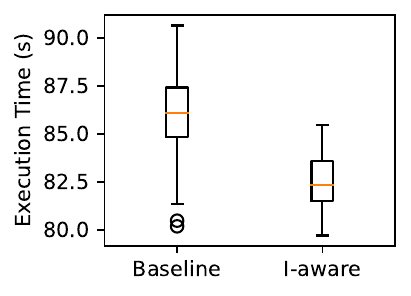}
        \caption{Hypre}\label{fig:corun_hypre}
    \end{subfigure}
    \begin{subfigure}[b]{0.325\linewidth}
        \centering
        \includegraphics[width=\textwidth]{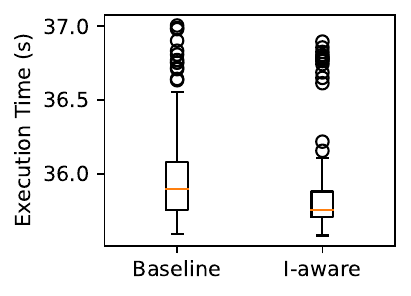}
        \caption{XSBench}\label{fig:corun_xs}
    \end{subfigure}
    \caption{The five-number summary of the execution time of each application in 100 runs with a mixture of co-running applications using a random baseline and interference-aware optimization.}
    %\caption{An interference-aware job scheduler can reduce workload execution time by avoiding co-locating sensitive applications on the same memory pool.}
    \label{fig:corun}
\end{figure}

The results in Section~\ref{sec:intf_sens} show that the performance impact of memory interference differs between different workloads. For practical adoption in HPC systems, job schedulers can be extended to take memory interference sensitivity into account for job co-location decisions. For instance, the user can use \benchname and the quantification method in Section~\ref{sec:interference} to quantify the application's sensitivity and provide it at job submission. Without interference awareness, interference-sensitive workloads may be scheduled together using the same memory pool, leading to poor performance.
%We propose to enhance the job scheduler by using sensitivity profiling data to avoid co-locating sensitive workloads on the same memory pool.
Interference-aware scheduling has shown benefit in previous works targeting other types of shared resources or configurations~\cite{delimitrou2013ibench,tzenetopoulos2022interference,zacarias2021intelligent,masouros2023adrias,google_pm,zacarias2021improving}. %An interference-aware scheduler can prevent interference-sensitive workloads from using the same memory pool as highly interference-inducing workloads.

To study the impact of interference-aware scheduling, we use the emulation setup and workloads from before with 50\% memory pool capacity. To simulate other jobs being scheduled onto the memory pool, we inject link traffic using \benchname. The level of interference changes randomly between 0--50\% every 60~s, representing different types of workloads. To simulate an interference-aware scheduler, which prevents interference-inducing jobs from being co-located, we vary the level of interference between 0--20\% instead. Each workload is run 100 times in both configurations.

The execution time in the baseline and interference-aware scheduler for each workload is shown in Figure~\ref{fig:corun}. In general, interference-aware scheduling improves execution time and reduces performance variability. The execution time is reduced because the average interference is lower with interference-aware scheduling. The variability is reduced because the range of possible interference levels is smaller with interference awareness as the high end of the spectrum is cut off. However, the results vary among the set of evaluated workloads. The average speedup is 4\% for Hypre, 2\% for NekRS and SuperLU, 1\% for BFS and HPL, and 0\% for XSBench. As a measure of variability, the 75th percentile of execution time decreased by 5\% for Hypre, 3\% for NekRS and SuperLU, 2\% BFS, and 1\% for HPL and XSBench.

As expected, these results largely match the measured interference coefficients for the workloads shown in Figure~\ref{fig:interference} and the interference sensitivity in Figure~\ref{fig:sensitivity50}. Hypre has the highest interference sensitivity and benefits most from interference-aware scheduling. Meanwhile, XSBench and HPL have the lowest interference sensitivity and show little benefit from interference-aware scheduling. The previous results showed that NekRS is more interference sensitive than BFS, but they show equal benefits from interference awareness.

These results indicate that integrating our quantitative methodology of memory interference into job schedulers such as SLURM could reduce interference and improve performance in a disaggregated memory system. The improvement of performance and reduced variation could incentivize users to inform their application's interference profile to the job scheduler and increase their confidence in the new memory architecture. Also, given that a 4\% performance improvement was observed for Hypre in our emulated environment, in an environment with higher induced interference, e.g., with more than two nodes per memory pool, the performance improvement could be more significant. We recognize that a real disaggregated system may have an even more skewed latency distribution than the one emulated in our study. Such a distribution has been found to have a high impact on some types of workloads and will be better studied using more accurate simulators such as gem5 and FPGA-based emulators~\cite{cilasun2022fpga,babaie2023enabling}. Nevertheless, our experiments with LBench still indirectly reflect the impact of random long latency induced by multiple jobs sharing a memory pool.
\section{Related Works}
In this section we summarize related works on multi-tier memory systems and disaggregated memory.

%\noindent
\textbf{Disaggregated Memory Systems.}
Both faulting methods (page-swapping) and non-faulting methods (direct cacheline access) have been proposed for implementing memory disaggregation~\cite{lim2009disaggregated}.
Until recently, network-based page-swapping has been the main practical option~\cite{lim2012system,gu2017efficient,bielski2016survey}. With major vendors adopting the CXL standard, many recent works focus on realizing non-faulting CXL-based memory disaggregation with a focus on cloud systems. 
\citeauthor{pinto2020thymesisflow} developed a full-stack prototype based on CXL-predecessor OpenCAPI using FPGAs, and \citeauthor{Gouk2022} developed a similar prototype for CXL~\cite{Gouk2022}. The implications of CXL-based memory for cloud workloads were studied in Meta datacenters~\cite{maruf2023tpp} and Microsoft Azure datacenters~\cite{pond23}. \citeauthor{lbl_dm} proposed an extended memory roofline model for designing HPC systems with disaggregated memory~\cite{lbl_dm}. They used analytical methods to estimate the local to remote memory access ratio in a set of AI trainings, data analytics workloads, and HPC micro-benchmarks. The access ratio was used in a roofline model to guide system design. \citeauthor{wahlgren2022evaluating} proposed an emulation method for CXL-based memory and focused on evaluating the performance penalty in HPC workloads~\cite{wahlgren2022evaluating}.
In contrast, our method is based on extensive quantitative measurements of memory access, application performance, and memory interference in HPC applications.

%\noindent
\textbf{Memory Interference and Contention.}
%The impact of memory interference on applications is gaining traction due to its impact on application. 
\citeauthor{tudor2011understanding} analyzed the impact of interference on parallel applications on multi-socket systems and developed an analytical model for memory contention based on the queuing theory~\cite{tudor2011understanding}. In infrastructures with workload co-location, prior works have investigated interference from shared hardware resources, such as L3 cache, network, etc, and their impact on performance and job scheduling policies~\cite{delimitrou2013ibench,tzenetopoulos2022interference,zacarias2021intelligent}. Recently, \citeauthor{masouros2023adrias} developed an interference-aware cloud scheduler for disaggregated memory systems~\cite{masouros2023adrias}, and \citeauthor{lee_mfence_2023} developed optimizations to mitigate interference for virtual machines with disaggregated memory~\cite{lee_mfence_2023}. \citeauthor{bsc_contention} developed a prediction model to estimate the performance degradation from interference in a disaggregated memory system~\cite{bsc_contention,zacarias2021improving}. They target a split architecture, where remote memory is provided by other compute nodes rather than dedicated memory servers, different from our target architecture in this work. 
%\citeauthor{google_pm} extended a cloud scheduler by using historical profiling data to reduce interference in a two-tiered system with persistent memory~\cite{google_pm}.
%\citeauthor{zacarias2021improving} later extended the prediction model to simulate a disaggregated memory cluster with a contention-aware SLURM job scheduler~\cite{zacarias2021improving}.
%Also, the emulation method and evaluation do not consider that the working set may be split between local and remote memory, i.e. the remote residency ratio is either 0\% or 100\%. 

%\noindent
\textbf{Data Placement in Multi-Tier Memory.}
%Most works on multi-tier memory systems aim to optimize data placement for cloud workloads. 
Extensive works have proposed automatic and transparent data placement solutions in multi-tier memory systems implemented with non-volatile memory or CXL. 
%For non-volatile memory, \citeauthor{dulloor2016data} developed X-Mem, an interface, profiler, and runtime for data placement~\cite{dulloor2016data}. 
\citeauthor{agarwal2017thermostat} proposed Thermostat, a fully application-transparent data placement runtime~\cite{agarwal2017thermostat} that detects hotness of pages for migration. \citeauthor{google_pm} evaluated a transparent runtime and optimized cluster scheduler in a production datacenter~\cite{google_pm}. For CXL-based memory, \citeauthor{maruf2023tpp} proposed TPP for transparent page placement for direct-attached memory~\cite{maruf2023tpp}, and \citeauthor{pond23} proposed Pond for data placement in shared memory pools~\cite{pond23}.

%While this approach is certainly useful for implementing disaggregated memory at a large scale, our case study shows that code changes guided by profiling can also benefit performance in a disaggregated setting. In particular, a runtime-based data placement system may incur overhead and increase unpredictability and may therefore be unsuitable for some HPC applications.
\section{Conclusions}
The recent adoption of disaggregated memory in several major data centers and the severe memory underutilization and high cost of memory in HPC facilities have mandated a closer look into the practical adoption of rack-scale memory pooling as a non-disruptive option on future HPC systems. This work described the prospects and requirements for adoption. In particular, we presented a multi-level quantitative methodology for dissecting application requirements on memory systems, from general to multi-tier, and memory pooling. We applied our method in NekRS, SuperLU, Hypre, HPL, BFS, and XSBench and identified key insights. In two case studies, we also demonstrated how findings from our method could be used for optimizing memory access at the application level and interference-aware job scheduling at the system level.
%While our current platform is limited to emulation on a single CPU node, we look forward to studying distributed workloads with production hardware and GPUs in the future. We hope this work can provide a foundation for further innovation in sustainable HPC architectures.

\section{Acknowledgments}
%Authors should not prepare this section using
%\begin{verbatim}
%  \begin{acks}
%  ...
%  \end{acks}
%\end{verbatim}

This research is supported by the European Commission under the Horizon project OpenCUBE (101092984) and the Swedish Research Council (no. 2022.03062). This work was partially performed under the auspices of the U.S. Department of Energy by Lawrence Livermore National Laboratory under contract No. DE-AC52-07NA27344 with support from the DOE Exascale Computing Project.

%%
%% The next two lines define the bibliography style to be used, and
%% the bibliography file.
\bibliographystyle{ACM-Reference-Format}
\bibliography{main,peng,jacob,gabin}

%%% -*-BibTeX-*-
%%% Do NOT edit. File created by BibTeX with style
%%% ACM-Reference-Format-Journals [18-Jan-2012].

\begin{thebibliography}{54}

%%% ====================================================================
%%% NOTE TO THE USER: you can override these defaults by providing
%%% customized versions of any of these macros before the \bibliography
%%% command.  Each of them MUST provide its own final punctuation,
%%% except for \shownote{}, \showDOI{}, and \showURL{}.  The latter two
%%% do not use final punctuation, in order to avoid confusing it with
%%% the Web address.
%%%
%%% To suppress output of a particular field, define its macro to expand
%%% to an empty string, or better, \unskip, like this:
%%%
%%% \newcommand{\showDOI}[1]{\unskip}   % LaTeX syntax
%%%
%%% \def \showDOI #1{\unskip}           % plain TeX syntax
%%%
%%% ====================================================================

\ifx \showCODEN    \undefined \def \showCODEN     #1{\unskip}     \fi
\ifx \showDOI      \undefined \def \showDOI       #1{#1}\fi
\ifx \showISBNx    \undefined \def \showISBNx     #1{\unskip}     \fi
\ifx \showISBNxiii \undefined \def \showISBNxiii  #1{\unskip}     \fi
\ifx \showISSN     \undefined \def \showISSN      #1{\unskip}     \fi
\ifx \showLCCN     \undefined \def \showLCCN      #1{\unskip}     \fi
\ifx \shownote     \undefined \def \shownote      #1{#1}          \fi
\ifx \showarticletitle \undefined \def \showarticletitle #1{#1}   \fi
\ifx \showURL      \undefined \def \showURL       {\relax}        \fi
% The following commands are used for tagged output and should be
% invisible to TeX
\providecommand\bibfield[2]{#2}
\providecommand\bibinfo[2]{#2}
\providecommand\natexlab[1]{#1}
\providecommand\showeprint[2][]{arXiv:#2}

\bibitem[Agarwal and Wenisch(2017)]%
        {agarwal2017thermostat}
\bibfield{author}{\bibinfo{person}{Neha Agarwal} {and}
  \bibinfo{person}{Thomas~F Wenisch}.} \bibinfo{year}{2017}\natexlab{}.
\newblock \showarticletitle{Thermostat: Application-transparent page management
  for two-tiered main memory}. In \bibinfo{booktitle}{\emph{Proceedings of the
  Twenty-Second International Conference on Architectural Support for
  Programming Languages and Operating Systems}}. \bibinfo{pages}{631--644}.
\newblock


\bibitem[Babaie et~al\mbox{.}(2023)]%
        {babaie2023enabling}
\bibfield{author}{\bibinfo{person}{Maryam Babaie}, \bibinfo{person}{Ayaz
  Akram}, {and} \bibinfo{person}{Jason Lowe-Power}.}
  \bibinfo{year}{2023}\natexlab{}.
\newblock \showarticletitle{Enabling Design Space Exploration of DRAM Caches
  for Emerging Memory Systems}. In \bibinfo{booktitle}{\emph{2023 IEEE
  International Symposium on Performance Analysis of Systems and Software
  (ISPASS)}}. IEEE, \bibinfo{pages}{340--342}.
\newblock


\bibitem[Bielski et~al\mbox{.}(2016)]%
        {bielski2016survey}
\bibfield{author}{\bibinfo{person}{Maciej Bielski}, \bibinfo{person}{Christian
  Pinto}, \bibinfo{person}{Daniel Raho}, {and} \bibinfo{person}{Renaud
  Pacalet}.} \bibinfo{year}{2016}\natexlab{}.
\newblock \showarticletitle{Survey on memory and devices disaggregation
  solutions for HPC systems}. In \bibinfo{booktitle}{\emph{2016 IEEE Intl
  Conference on Computational Science and Engineering (CSE) and IEEE Intl
  Conference on Embedded and Ubiquitous Computing (EUC) and 15th Intl Symposium
  on Distributed Computing and Applications for Business Engineering
  (DCABES)}}. IEEE, \bibinfo{pages}{197--204}.
\newblock


\bibitem[Cineca(2022)]%
        {leonardoSpecs}
\bibfield{author}{\bibinfo{person}{Cineca}.} \bibinfo{year}{2022}\natexlab{}.
\newblock \bibinfo{title}{{Leonardo HPC System}}.
\newblock
  \bibinfo{howpublished}{\url{https://leonardo-supercomputer.cineca.eu/hpc-system/}}.
\newblock


\bibitem[Cılasun et~al\mbox{.}(2022)]%
        {cilasun2022fpga}
\bibfield{author}{\bibinfo{person}{Hüsrev Cılasun},
  \bibinfo{person}{Christopher Macaraeg}, \bibinfo{person}{Ivy Peng},
  \bibinfo{person}{Abhik Sarkar}, {and} \bibinfo{person}{Maya Gokhale}.}
  \bibinfo{year}{2022}\natexlab{}.
\newblock \showarticletitle{{FPGA}-accelerated simulation of variable latency
  memory systems}. In \bibinfo{booktitle}{\emph{Proceedings of the
  International Symposium on Memory Systems}} \emph{(\bibinfo{series}{MEMSYS
  '22})}.
\newblock


\bibitem[Davis and Hu(2011)]%
        {Davis2011}
\bibfield{author}{\bibinfo{person}{Timothy~A. Davis} {and}
  \bibinfo{person}{Yifan Hu}.} \bibinfo{year}{2011}\natexlab{}.
\newblock \showarticletitle{The university of {Florida} sparse matrix
  collection}.
\newblock \bibinfo{journal}{\emph{ACM Trans. Math. Software}}
  \bibinfo{volume}{38}, \bibinfo{number}{1} (\bibinfo{date}{nov}
  \bibinfo{year}{2011}), \bibinfo{pages}{1--25}.
\newblock
\urldef\tempurl%
\url{https://doi.org/10.1145/2049662.2049663}
\showDOI{\tempurl}


\bibitem[Delimitrou and Kozyrakis(2013)]%
        {delimitrou2013ibench}
\bibfield{author}{\bibinfo{person}{Christina Delimitrou} {and}
  \bibinfo{person}{Christos Kozyrakis}.} \bibinfo{year}{2013}\natexlab{}.
\newblock \showarticletitle{ibench: Quantifying interference for datacenter
  applications}. In \bibinfo{booktitle}{\emph{2013 IEEE international symposium
  on workload characterization (IISWC)}}. IEEE, \bibinfo{pages}{23--33}.
\newblock


\bibitem[Ding et~al\mbox{.}(2022)]%
        {lbl_dm}
\bibfield{author}{\bibinfo{person}{Nan Ding}, \bibinfo{person}{Samuel
  Williams}, \bibinfo{person}{Hai~Ah Nam}, \bibinfo{person}{Taylor Groves},
  \bibinfo{person}{Muaaz~Gul Awan}, \bibinfo{person}{LeAnn Lindsey},
  \bibinfo{person}{Christopher Daley}, \bibinfo{person}{Oguz Selvitopi},
  \bibinfo{person}{Leonid Oliker}, {and} \bibinfo{person}{Nicholas Wright}.}
  \bibinfo{year}{2022}\natexlab{}.
\newblock \showarticletitle{Methodology for Evaluating the Potential of
  Disaggregated Memory Systems}. In \bibinfo{booktitle}{\emph{2022 IEEE/ACM
  International Workshop on Resource Disaggregation in High-Performance
  Computing (REDIS)}}. \bibinfo{pages}{1--11}.
\newblock


\bibitem[Dongarra(2016)]%
        {sunwaySpecsReport}
\bibfield{author}{\bibinfo{person}{Jack Dongarra}.}
  \bibinfo{year}{2016}\natexlab{}.
\newblock \bibinfo{title}{{Report on the Sunway TaihuLight System}}.
\newblock
  \bibinfo{howpublished}{\url{https://netlib.org/utk/people/JackDongarra/PAPERS/sunway-report-2016.pdf}}.
\newblock


\bibitem[Dongarra(2017)]%
        {tianheSpecs}
\bibfield{author}{\bibinfo{person}{Jack Dongarra}.}
  \bibinfo{year}{2017}\natexlab{}.
\newblock \bibinfo{title}{{Report on the Tianhe-2A System}}.
\newblock
  \bibinfo{howpublished}{\url{https://icl.utk.edu/files/publications/2017/icl-utk-970-2017.pdf}}.
\newblock


\bibitem[Dongarra et~al\mbox{.}(2003)]%
        {dongarra2003linpack}
\bibfield{author}{\bibinfo{person}{Jack~J Dongarra}, \bibinfo{person}{Piotr
  Luszczek}, {and} \bibinfo{person}{Antoine Petitet}.}
  \bibinfo{year}{2003}\natexlab{}.
\newblock \showarticletitle{The LINPACK benchmark: past, present and future}.
\newblock \bibinfo{journal}{\emph{Concurrency and Computation: practice and
  experience}} \bibinfo{volume}{15}, \bibinfo{number}{9}
  (\bibinfo{year}{2003}), \bibinfo{pages}{803--820}.
\newblock


\bibitem[Duraisamy et~al\mbox{.}(2023)]%
        {google_pm}
\bibfield{author}{\bibinfo{person}{Padmapriya Duraisamy}, \bibinfo{person}{Wei
  Xu}, \bibinfo{person}{Scott Hare}, \bibinfo{person}{Ravi Rajwar},
  \bibinfo{person}{David Culler}, \bibinfo{person}{Zhiyi Xu},
  \bibinfo{person}{Jianing Fan}, \bibinfo{person}{Christopher Kennelly},
  \bibinfo{person}{Bill McCloskey}, \bibinfo{person}{Danijela Mijailovic},
  \bibinfo{person}{Brian Morris}, \bibinfo{person}{Chiranjit Mukherjee},
  \bibinfo{person}{Jingliang Ren}, \bibinfo{person}{Greg Thelen},
  \bibinfo{person}{Paul Turner}, \bibinfo{person}{Carlos Villavieja},
  \bibinfo{person}{Parthasarathy Ranganathan}, {and} \bibinfo{person}{Amin
  Vahdat}.} \bibinfo{year}{2023}\natexlab{}.
\newblock \showarticletitle{Towards an Adaptable Systems Architecture for
  Memory Tiering at Warehouse-Scale}. In \bibinfo{booktitle}{\emph{Proceedings
  of the 28th ACM International Conference on Architectural Support for
  Programming Languages and Operating Systems, Volume 3}} (Vancouver, BC,
  Canada) \emph{(\bibinfo{series}{ASPLOS 2023})}.
  \bibinfo{publisher}{Association for Computing Machinery},
  \bibinfo{address}{New York, NY, USA}, \bibinfo{pages}{727–741}.
\newblock
\showISBNx{9781450399180}
\urldef\tempurl%
\url{https://doi.org/10.1145/3582016.3582031}
\showDOI{\tempurl}


\bibitem[Eun-jin(2023)]%
        {memoryCost2023}
\bibfield{author}{\bibinfo{person}{By~Kim Eun-jin}.}
  \bibinfo{year}{2023}\natexlab{}.
\newblock \showarticletitle{Samsung and SK Hynix Enjoy a Rush of Orders for New
  Memories}.
\newblock
  \bibinfo{howpublished}{\url{http://www.businesskorea.co.kr/news/articleView.html?idxno=109380}}.
\newblock \bibinfo{journal}{\emph{Business Korea}} (\bibinfo{year}{2023}).
\newblock


\bibitem[Ewais and Chow(2023)]%
        {ewais_disaggregated_2023}
\bibfield{author}{\bibinfo{person}{Mohammad Ewais} {and} \bibinfo{person}{Paul
  Chow}.} \bibinfo{year}{2023}\natexlab{}.
\newblock \showarticletitle{Disaggregated {Memory} in the {Datacenter}: {A}
  {Survey}}.
\newblock \bibinfo{journal}{\emph{IEEE Access}} (\bibinfo{year}{2023}).
\newblock
\showISSN{2169-3536}
\newblock
\shownote{Publisher: IEEE}.


\bibitem[Falgout and Yang(2002)]%
        {hypre2}
\bibfield{author}{\bibinfo{person}{Robert~D Falgout} {and}
  \bibinfo{person}{Ulrike~Meier Yang}.} \bibinfo{year}{2002}\natexlab{}.
\newblock \showarticletitle{hypre: A library of high performance
  preconditioners}. In \bibinfo{booktitle}{\emph{Computational Science—ICCS
  2002: International Conference Amsterdam, The Netherlands, April 21--24, 2002
  Proceedings, Part III}}. Springer, \bibinfo{pages}{632--641}.
\newblock


\bibitem[Fischer et~al\mbox{.}(2022)]%
        {fischer2021nekrs}
\bibfield{author}{\bibinfo{person}{Paul Fischer}, \bibinfo{person}{Stefan
  Kerkemeier}, \bibinfo{person}{Misun Min}, \bibinfo{person}{Yu-Hsiang Lan},
  \bibinfo{person}{Malachi Phillips}, \bibinfo{person}{Thilina Rathnayake},
  \bibinfo{person}{Elia Merzari}, \bibinfo{person}{Ananias Tomboulides},
  \bibinfo{person}{Ali Karakus}, \bibinfo{person}{Noel Chalmers},
  {et~al\mbox{.}}} \bibinfo{year}{2022}\natexlab{}.
\newblock \showarticletitle{NekRS, a GPU-accelerated spectral element
  Navier--Stokes solver}.
\newblock \bibinfo{journal}{\emph{Parallel Comput.}}  \bibinfo{volume}{114}
  (\bibinfo{year}{2022}), \bibinfo{pages}{102982}.
\newblock


\bibitem[Fujitsu(2020)]%
        {fugakuSpecs}
\bibfield{author}{\bibinfo{person}{Fujitsu}.} \bibinfo{year}{2020}\natexlab{}.
\newblock \bibinfo{title}{{Supercomputer Fugaku -- Specifications}}.
\newblock
  \bibinfo{howpublished}{\url{https://www.fujitsu.com/global/about/innovation/fugaku/specifications/}}.
\newblock


\bibitem[Gouk et~al\mbox{.}(2022)]%
        {Gouk2022}
\bibfield{author}{\bibinfo{person}{Donghyun Gouk}, \bibinfo{person}{Sangwon
  Lee}, \bibinfo{person}{Miryeong Kwon}, {and} \bibinfo{person}{Myoungsoo
  Jung}.} \bibinfo{year}{2022}\natexlab{}.
\newblock \showarticletitle{Direct Access, High-Performance Memory
  Disaggregation with {DirectCXL}}. In \bibinfo{booktitle}{\emph{2022 USENIX
  Annual Technical Conference (USENIX ATC 22)}}. \bibinfo{pages}{287--294}.
\newblock


\bibitem[Gu et~al\mbox{.}(2017)]%
        {gu2017efficient}
\bibfield{author}{\bibinfo{person}{Juncheng Gu}, \bibinfo{person}{Youngmoon
  Lee}, \bibinfo{person}{Yiwen Zhang}, \bibinfo{person}{Mosharaf Chowdhury},
  {and} \bibinfo{person}{Kang~G Shin}.} \bibinfo{year}{2017}\natexlab{}.
\newblock \showarticletitle{Efficient Memory Disaggregation with Infiniswap.}.
  In \bibinfo{booktitle}{\emph{NSDI}}. \bibinfo{pages}{649--667}.
\newblock


\bibitem[HPCWire(2021)]%
        {llnlrabbit21}
\bibfield{author}{\bibinfo{person}{HPCWire}.} \bibinfo{year}{2021}\natexlab{}.
\newblock \bibinfo{title}{{Livermore’s El Capitan Supercomputer to Debut HPE
  ‘Rabbit’ Near Node Local Storage}}.
\newblock
  \bibinfo{howpublished}{\url{https://hpc.llnl.gov/livermore’s-el-capitan-supercomputer-debut-hpe-rabbit-‘near-node’-storage}}.
\newblock


\bibitem[Laboratory(2018)]%
        {sierraSpecs}
\bibfield{author}{\bibinfo{person}{Lawrence Livermore~National Laboratory}.}
  \bibinfo{year}{2018}\natexlab{}.
\newblock \bibinfo{title}{{Sierra}}.
\newblock
  \bibinfo{howpublished}{\url{https://hpc.llnl.gov/hardware/compute-platforms/sierra}}.
\newblock


\bibitem[Laboratory(2021)]%
        {frontierSpecs}
\bibfield{author}{\bibinfo{person}{Oak Ridge~National Laboratory}.}
  \bibinfo{year}{2021}\natexlab{}.
\newblock \bibinfo{title}{{Frontier User Guide}}.
\newblock
  \bibinfo{howpublished}{\url{https://docs.olcf.ornl.gov/systems/frontier_user_guide.html}}.
\newblock


\bibitem[Lee et~al\mbox{.}(2023)]%
        {lee_mfence_2023}
\bibfield{author}{\bibinfo{person}{Jinhoon Lee}, \bibinfo{person}{Yeonwoo
  Jung}, \bibinfo{person}{Suyeon Lee}, \bibinfo{person}{Safdar Jamil},
  \bibinfo{person}{Sungyong Park}, \bibinfo{person}{Kwangwon Koh},
  \bibinfo{person}{Hongyeon Kim}, \bibinfo{person}{Kangho Kim}, {and}
  \bibinfo{person}{Youngjae Kim}.} \bibinfo{year}{2023}\natexlab{}.
\newblock \showarticletitle{{MFence}: {Defending} {Against} {Memory} {Access}
  {Interference} in a {Disaggregated} {Cloud} {Memory} {Platform}}.
\newblock  (\bibinfo{year}{2023}).
\newblock


\bibitem[Li et~al\mbox{.}(2023)]%
        {pond23}
\bibfield{author}{\bibinfo{person}{Huaicheng Li}, \bibinfo{person}{Daniel~S.
  Berger}, \bibinfo{person}{Lisa Hsu}, \bibinfo{person}{Daniel Ernst},
  \bibinfo{person}{Pantea Zardoshti}, \bibinfo{person}{Stanko Novakovic},
  \bibinfo{person}{Monish Shah}, \bibinfo{person}{Samir Rajadnya},
  \bibinfo{person}{Scott Lee}, \bibinfo{person}{Ishwar Agarwal},
  \bibinfo{person}{Mark~D. Hill}, \bibinfo{person}{Marcus Fontoura}, {and}
  \bibinfo{person}{Ricardo Bianchini}.} \bibinfo{year}{2023}\natexlab{}.
\newblock \showarticletitle{Pond: CXL-Based Memory Pooling Systems for Cloud
  Platforms}. In \bibinfo{booktitle}{\emph{Proceedings of the 28th ACM
  International Conference on Architectural Support for Programming Languages
  and Operating Systems, Volume 2}} (Vancouver, BC, Canada)
  \emph{(\bibinfo{series}{ASPLOS 2023})}. \bibinfo{publisher}{Association for
  Computing Machinery}, \bibinfo{address}{New York, NY, USA},
  \bibinfo{pages}{574–587}.
\newblock
\showISBNx{9781450399166}
\urldef\tempurl%
\url{https://doi.org/10.1145/3575693.3578835}
\showDOI{\tempurl}


\bibitem[Li(2005)]%
        {Li2005}
\bibfield{author}{\bibinfo{person}{Xiaoye~S. Li}.}
  \bibinfo{year}{2005}\natexlab{}.
\newblock \showarticletitle{An Overview of {SuperLU}: Algorithms,
  Implementation, and User Interface}.
\newblock \bibinfo{journal}{\emph{ACM Trans. Math. Softw.}}
  \bibinfo{volume}{31}, \bibinfo{number}{3} (\bibinfo{date}{sep}
  \bibinfo{year}{2005}), \bibinfo{pages}{302–325}.
\newblock
\showISSN{0098-3500}
\urldef\tempurl%
\url{https://doi.org/10.1145/1089014.1089017}
\showDOI{\tempurl}


\bibitem[Lim et~al\mbox{.}(2009)]%
        {lim2009disaggregated}
\bibfield{author}{\bibinfo{person}{Kevin Lim}, \bibinfo{person}{Jichuan Chang},
  \bibinfo{person}{Trevor Mudge}, \bibinfo{person}{Parthasarathy Ranganathan},
  \bibinfo{person}{Steven~K Reinhardt}, {and} \bibinfo{person}{Thomas~F
  Wenisch}.} \bibinfo{year}{2009}\natexlab{}.
\newblock \showarticletitle{Disaggregated memory for expansion and sharing in
  blade servers}.
\newblock \bibinfo{journal}{\emph{ACM SIGARCH computer architecture news}}
  \bibinfo{volume}{37}, \bibinfo{number}{3} (\bibinfo{year}{2009}),
  \bibinfo{pages}{267--278}.
\newblock


\bibitem[Lim et~al\mbox{.}(2012)]%
        {lim2012system}
\bibfield{author}{\bibinfo{person}{Kevin Lim}, \bibinfo{person}{Yoshio Turner},
  \bibinfo{person}{Jose~Renato Santos}, \bibinfo{person}{Alvin AuYoung},
  \bibinfo{person}{Jichuan Chang}, \bibinfo{person}{Parthasarathy Ranganathan},
  {and} \bibinfo{person}{Thomas~F Wenisch}.} \bibinfo{year}{2012}\natexlab{}.
\newblock \showarticletitle{System-level implications of disaggregated memory}.
  In \bibinfo{booktitle}{\emph{IEEE International Symposium on High-Performance
  Comp Architecture}}. IEEE, \bibinfo{pages}{1--12}.
\newblock


\bibitem[LUMI(2022)]%
        {lumiSpecs}
\bibfield{author}{\bibinfo{person}{LUMI}.} \bibinfo{year}{2022}\natexlab{}.
\newblock \bibinfo{title}{{LUMI Documentation -- Hardware}}.
\newblock
  \bibinfo{howpublished}{\url{https://docs.lumi-supercomputer.eu/hardware/compute/lumig/}}.
\newblock


\bibitem[Mahar et~al\mbox{.}(2023)]%
        {mahar2023workload}
\bibfield{author}{\bibinfo{person}{Suyash Mahar}, \bibinfo{person}{Hao Wang},
  \bibinfo{person}{Wei Shu}, {and} \bibinfo{person}{Abhishek Dhanotia}.}
  \bibinfo{year}{2023}\natexlab{}.
\newblock \showarticletitle{Workload Behavior Driven Memory Subsystem Design
  for Hyperscale}.
\newblock \bibinfo{journal}{\emph{arXiv preprint arXiv:2303.08396}}
  (\bibinfo{year}{2023}).
\newblock


\bibitem[Maruf et~al\mbox{.}(2023)]%
        {maruf2023tpp}
\bibfield{author}{\bibinfo{person}{Hasan~Al Maruf}, \bibinfo{person}{Hao Wang},
  \bibinfo{person}{Abhishek Dhanotia}, \bibinfo{person}{Johannes Weiner},
  \bibinfo{person}{Niket Agarwal}, \bibinfo{person}{Pallab Bhattacharya},
  \bibinfo{person}{Chris Petersen}, \bibinfo{person}{Mosharaf Chowdhury},
  \bibinfo{person}{Shobhit Kanaujia}, {and} \bibinfo{person}{Prakash Chauhan}.}
  \bibinfo{year}{2023}\natexlab{}.
\newblock \showarticletitle{TPP: Transparent Page Placement for CXL-Enabled
  Tiered-Memory}. In \bibinfo{booktitle}{\emph{Proceedings of the 28th ACM
  International Conference on Architectural Support for Programming Languages
  and Operating Systems, Volume 3}}. \bibinfo{pages}{742--755}.
\newblock


\bibitem[Masouros et~al\mbox{.}(2023)]%
        {masouros2023adrias}
\bibfield{author}{\bibinfo{person}{Dimosthenis Masouros},
  \bibinfo{person}{Christian Pinto}, \bibinfo{person}{Michele Gazzetti},
  \bibinfo{person}{Sotirios Xydis}, {and} \bibinfo{person}{Dimitrios Soudris}.}
  \bibinfo{year}{2023}\natexlab{}.
\newblock \showarticletitle{Adrias: Interference-Aware Memory Orchestration for
  Disaggregated Cloud Infrastructures}. In \bibinfo{booktitle}{\emph{2023 IEEE
  International Symposium on High-Performance Computer Architecture (HPCA)}}.
  IEEE, \bibinfo{pages}{855--869}.
\newblock


\bibitem[Matsuoka et~al\mbox{.}(2023)]%
        {matsuoka2023myths}
\bibfield{author}{\bibinfo{person}{Satoshi Matsuoka}, \bibinfo{person}{Jens
  Domke}, \bibinfo{person}{Mohamed Wahib}, \bibinfo{person}{Aleksandr Drozd},
  {and} \bibinfo{person}{Torsten Hoefler}.} \bibinfo{year}{2023}\natexlab{}.
\newblock \showarticletitle{Myths and Legends in High-Performance Computing}.
\newblock \bibinfo{journal}{\emph{arXiv preprint arXiv:2301.02432}}
  (\bibinfo{year}{2023}).
\newblock


\bibitem[Michelogiannakis et~al\mbox{.}(2022)]%
        {michelogiannakis2022case}
\bibfield{author}{\bibinfo{person}{George Michelogiannakis},
  \bibinfo{person}{Benjamin Klenk}, \bibinfo{person}{Brandon Cook},
  \bibinfo{person}{Min~Yee Teh}, \bibinfo{person}{Madeleine Glick},
  \bibinfo{person}{Larry Dennison}, \bibinfo{person}{Keren Bergman}, {and}
  \bibinfo{person}{John Shalf}.} \bibinfo{year}{2022}\natexlab{}.
\newblock \showarticletitle{A case for intra-rack resource disaggregation in
  hpc}.
\newblock \bibinfo{journal}{\emph{ACM Transactions on Architecture and Code
  Optimization (TACO)}} \bibinfo{volume}{19}, \bibinfo{number}{2}
  (\bibinfo{year}{2022}), \bibinfo{pages}{1--26}.
\newblock


\bibitem[(NERSC)(2021)]%
        {perlmutterSpecs}
\bibfield{author}{\bibinfo{person}{National Energy Research
  Scientific~Computing (NERSC)}.} \bibinfo{year}{2021}\natexlab{}.
\newblock \bibinfo{title}{{Perlmutter Architecture}}.
\newblock
  \bibinfo{howpublished}{\url{https://docs.nersc.gov/systems/perlmutter/architecture/}}.
\newblock


\bibitem[NVIDIA(2020)]%
        {seleneSpecs}
\bibfield{author}{\bibinfo{person}{NVIDIA}.} \bibinfo{year}{2020}\natexlab{}.
\newblock \bibinfo{title}{{NVIDIA Selene: Leadership-Class Supercomputing
  Infrastructure}}.
\newblock
  \bibinfo{howpublished}{\url{https://www.nvidia.com/en-us/on-demand/session/supercomputing2020-sc2019/}}.
\newblock


\bibitem[Peng et~al\mbox{.}(2021)]%
        {peng2021holistic}
\bibfield{author}{\bibinfo{person}{Ivy Peng}, \bibinfo{person}{Ian Karlin},
  \bibinfo{person}{Maya Gokhale}, \bibinfo{person}{Kathleen Shoga},
  \bibinfo{person}{Matthew Legendre}, {and} \bibinfo{person}{Todd Gamblin}.}
  \bibinfo{year}{2021}\natexlab{}.
\newblock \showarticletitle{A holistic view of memory utilization on HPC
  systems: Current and future trends}. In \bibinfo{booktitle}{\emph{The
  International Symposium on Memory Systems}}. \bibinfo{pages}{1--11}.
\newblock


\bibitem[Peng et~al\mbox{.}(2020a)]%
        {peng2020memory}
\bibfield{author}{\bibinfo{person}{Ivy Peng}, \bibinfo{person}{Roger Pearce},
  {and} \bibinfo{person}{Maya Gokhale}.} \bibinfo{year}{2020}\natexlab{a}.
\newblock \showarticletitle{On the memory underutilization: Exploring
  disaggregated memory on hpc systems}. In \bibinfo{booktitle}{\emph{2020 IEEE
  32nd International Symposium on Computer Architecture and High Performance
  Computing (SBAC-PAD)}}. IEEE, \bibinfo{pages}{183--190}.
\newblock


\bibitem[Peng et~al\mbox{.}(2020b)]%
        {peng2020demystifying}
\bibfield{author}{\bibinfo{person}{Ivy Peng}, \bibinfo{person}{Kai Wu},
  \bibinfo{person}{Jie Ren}, \bibinfo{person}{Dong Li}, {and}
  \bibinfo{person}{Maya Gokhale}.} \bibinfo{year}{2020}\natexlab{b}.
\newblock \showarticletitle{Demystifying the performance of hpc scientific
  applications on nvm-based memory systems}. In \bibinfo{booktitle}{\emph{2020
  IEEE International Parallel and Distributed Processing Symposium (IPDPS)}}.
  IEEE, \bibinfo{pages}{916--925}.
\newblock


\bibitem[Pinto et~al\mbox{.}(2020)]%
        {pinto2020thymesisflow}
\bibfield{author}{\bibinfo{person}{Christian Pinto}, \bibinfo{person}{Dimitris
  Syrivelis}, \bibinfo{person}{Michele Gazzetti}, \bibinfo{person}{Panos
  Koutsovasilis}, \bibinfo{person}{Andrea Reale}, \bibinfo{person}{Kostas
  Katrinis}, {and} \bibinfo{person}{H~Peter Hofstee}.}
  \bibinfo{year}{2020}\natexlab{}.
\newblock \showarticletitle{Thymesisflow: A software-defined, hw/sw co-designed
  interconnect stack for rack-scale memory disaggregation}. In
  \bibinfo{booktitle}{\emph{2020 53rd Annual IEEE/ACM International Symposium
  on Microarchitecture (MICRO)}}. IEEE, \bibinfo{pages}{868--880}.
\newblock


\bibitem[Shun and Blelloch(2013)]%
        {shun2013ligra}
\bibfield{author}{\bibinfo{person}{Julian Shun} {and} \bibinfo{person}{Guy~E.
  Blelloch}.} \bibinfo{year}{2013}\natexlab{}.
\newblock \showarticletitle{Ligra: a lightweight graph processing framework for
  shared memory}.
\newblock \bibinfo{journal}{\emph{{ACM} {SIGPLAN} Notices}}
  \bibinfo{volume}{48}, \bibinfo{number}{8} (\bibinfo{date}{aug}
  \bibinfo{year}{2013}), \bibinfo{pages}{135--146}.
\newblock
\urldef\tempurl%
\url{https://doi.org/10.1145/2517327.2442530}
\showDOI{\tempurl}


\bibitem[Sun et~al\mbox{.}(2023)]%
        {sun2023demystifying}
\bibfield{author}{\bibinfo{person}{Yan Sun}, \bibinfo{person}{Yifan Yuan},
  \bibinfo{person}{Zeduo Yu}, \bibinfo{person}{Reese Kuper},
  \bibinfo{person}{Ipoom Jeong}, \bibinfo{person}{Ren Wang}, {and}
  \bibinfo{person}{Nam~Sung Kim}.} \bibinfo{year}{2023}\natexlab{}.
\newblock \showarticletitle{Demystifying CXL Memory with Genuine CXL-Ready
  Systems and Devices}.
\newblock \bibinfo{journal}{\emph{arXiv preprint arXiv:2303.15375}}
  (\bibinfo{year}{2023}).
\newblock


\bibitem[{TOP500.org}(2022)]%
        {top500}
\bibfield{author}{\bibinfo{person}{{TOP500.org}}.}
  \bibinfo{year}{2022}\natexlab{}.
\newblock \bibinfo{title}{Top500 List November 2022}.
\newblock
\newblock
\urldef\tempurl%
\url{https://www.top500.org/lists/top500/2022/11/}
\showURL{%
\tempurl}


\bibitem[Tramm et~al\mbox{.}(2014)]%
        {tramm2014xsbench}
\bibfield{author}{\bibinfo{person}{John~R Tramm}, \bibinfo{person}{Andrew~R
  Siegel}, \bibinfo{person}{Tanzima Islam}, {and} \bibinfo{person}{Martin
  Schulz}.} \bibinfo{year}{2014}\natexlab{}.
\newblock \showarticletitle{{XSBench} - The Development and Verification of a
  Performance Abstraction for {M}onte {C}arlo Reactor Analysis}. In
  \bibinfo{booktitle}{\emph{{PHYSOR} 2014 - The Role of Reactor Physics toward
  a Sustainable Future}}. \bibinfo{address}{Kyoto}.
\newblock
\urldef\tempurl%
\url{https://www.mcs.anl.gov/papers/P5064-0114.pdf}
\showURL{%
\tempurl}


\bibitem[Tsai et~al\mbox{.}(2020)]%
        {tsai2020disaggregating}
\bibfield{author}{\bibinfo{person}{Shin-Yeh Tsai}, \bibinfo{person}{Yizhou
  Shan}, {and} \bibinfo{person}{Yiying Zhang}.}
  \bibinfo{year}{2020}\natexlab{}.
\newblock \showarticletitle{Disaggregating persistent memory and controlling
  them remotely: An exploration of passive disaggregated key-value stores}. In
  \bibinfo{booktitle}{\emph{Proceedings of the 2020 USENIX Conference on Usenix
  Annual Technical Conference}}. \bibinfo{pages}{33--48}.
\newblock


\bibitem[Tudor et~al\mbox{.}(2011)]%
        {tudor2011understanding}
\bibfield{author}{\bibinfo{person}{Bogdan~Marius Tudor},
  \bibinfo{person}{Yong~Meng Teo}, {and} \bibinfo{person}{Simon See}.}
  \bibinfo{year}{2011}\natexlab{}.
\newblock \showarticletitle{Understanding off-chip memory contention of
  parallel programs in multicore systems}. In \bibinfo{booktitle}{\emph{2011
  International Conference on Parallel Processing}}. IEEE,
  \bibinfo{pages}{602--611}.
\newblock


\bibitem[Tzenetopoulos et~al\mbox{.}(2022)]%
        {tzenetopoulos2022interference}
\bibfield{author}{\bibinfo{person}{Achilleas Tzenetopoulos},
  \bibinfo{person}{Dimosthenis Masouros}, \bibinfo{person}{Sotirios Xydis},
  {and} \bibinfo{person}{Dimitrios Soudris}.} \bibinfo{year}{2022}\natexlab{}.
\newblock \showarticletitle{Interference-aware workload placement for improving
  latency distribution of converged HPC/Big Data cloud infrastructures}. In
  \bibinfo{booktitle}{\emph{Embedded Computer Systems: Architectures, Modeling,
  and Simulation: 21st International Conference, SAMOS 2021, Virtual Event,
  July 4--8, 2021, Proceedings}}. Springer, \bibinfo{pages}{108--123}.
\newblock


\bibitem[user Guide(2018)]%
        {summitSpecs}
\bibfield{author}{\bibinfo{person}{Summit user Guide}.}
  \bibinfo{year}{2018}\natexlab{}.
\newblock \bibinfo{title}{{Summit}}.
\newblock
  \bibinfo{howpublished}{\url{https://docs.olcf.ornl.gov/systems/summit_user_guide.html}}.
\newblock


\bibitem[Wahlgren et~al\mbox{.}(2022)]%
        {wahlgren2022evaluating}
\bibfield{author}{\bibinfo{person}{Jacob Wahlgren}, \bibinfo{person}{Maya
  Gokhale}, {and} \bibinfo{person}{Ivy~B Peng}.}
  \bibinfo{year}{2022}\natexlab{}.
\newblock \showarticletitle{Evaluating Emerging CXL-enabled Memory Pooling for
  HPC Systems}.
\newblock \bibinfo{journal}{\emph{2022 IEEE/ACM Workshop on Memory Centric High
  Performance Computing}} (\bibinfo{year}{2022}).
\newblock


\bibitem[Wang et~al\mbox{.}(2023)]%
        {wangcanvas2022}
\bibfield{author}{\bibinfo{person}{Chenxi Wang}, \bibinfo{person}{Yifan Qiao},
  \bibinfo{person}{Haoran Ma}, \bibinfo{person}{Shi Liu},
  \bibinfo{person}{Yiying Zhang}, \bibinfo{person}{Wenguang Chen},
  \bibinfo{person}{Ravi Netravali}, \bibinfo{person}{Miryung Kim}, {and}
  \bibinfo{person}{Guoqing~Harry Xu}.} \bibinfo{year}{2023}\natexlab{}.
\newblock \showarticletitle{Canvas: {Isolated} and adaptive swapping for
  multi-applications on remote memory}. In \bibinfo{booktitle}{\emph{NSDI}}.
\newblock


\bibitem[Weiner(2022)]%
        {weiner_2022}
\bibfield{author}{\bibinfo{person}{Johannes Weiner}.}
  \bibinfo{year}{2022}\natexlab{}.
\newblock \bibinfo{title}{[PATCH] mm: mempolicy: N:M interleave policy for
  tiered memory nodes}.
\newblock
\newblock
\urldef\tempurl%
\url{https://lore.kernel.org/linux-mm/YqD0\%2FtzFwXvJ1gK6@cmpxchg.org/T/}
\showURL{%
\tempurl}


\bibitem[Williams et~al\mbox{.}(2009)]%
        {williams_roofline_2009}
\bibfield{author}{\bibinfo{person}{Samuel Williams}, \bibinfo{person}{Andrew
  Waterman}, {and} \bibinfo{person}{David Patterson}.}
  \bibinfo{year}{2009}\natexlab{}.
\newblock \showarticletitle{Roofline: an insightful visual performance model
  for multicore architectures}.
\newblock \bibinfo{journal}{\emph{Commun. ACM}} \bibinfo{volume}{52},
  \bibinfo{number}{4} (\bibinfo{year}{2009}), \bibinfo{pages}{65--76}.
\newblock
\showISSN{0001-0782}
\newblock
\shownote{Publisher: ACM New York, NY, USA}.


\bibitem[Zacarias et~al\mbox{.}(2021a)]%
        {zacarias2021improving}
\bibfield{author}{\bibinfo{person}{Felippe~Vieira Zacarias},
  \bibinfo{person}{Paul Carpenter}, {and} \bibinfo{person}{Vinicius Petrucci}.}
  \bibinfo{year}{2021}\natexlab{a}.
\newblock \showarticletitle{Improving hpc system throughput and response time
  using memory disaggregation}. In \bibinfo{booktitle}{\emph{2021 IEEE 27th
  International Conference on Parallel and Distributed Systems (ICPADS)}}.
  IEEE, \bibinfo{pages}{283--290}.
\newblock


\bibitem[Zacarias et~al\mbox{.}(2020)]%
        {bsc_contention}
\bibfield{author}{\bibinfo{person}{Felippe~Vieira Zacarias},
  \bibinfo{person}{Rajiv Nishtala}, {and} \bibinfo{person}{Paul Carpenter}.}
  \bibinfo{year}{2020}\natexlab{}.
\newblock \showarticletitle{Contention-aware application performance prediction
  for disaggregated memory systems}. In \bibinfo{booktitle}{\emph{Proceedings
  of the 17th ACM International Conference on Computing Frontiers}}.
  \bibinfo{pages}{49--59}.
\newblock


\bibitem[Zacarias et~al\mbox{.}(2021b)]%
        {zacarias2021intelligent}
\bibfield{author}{\bibinfo{person}{Felippe~Vieira Zacarias},
  \bibinfo{person}{Vinicius Petrucci}, \bibinfo{person}{Rajiv Nishtala},
  \bibinfo{person}{Paul Carpenter}, {and} \bibinfo{person}{Daniel Moss{\'e}}.}
  \bibinfo{year}{2021}\natexlab{b}.
\newblock \showarticletitle{Intelligent colocation of HPC workloads}.
\newblock \bibinfo{journal}{\emph{J. Parallel and Distrib. Comput.}}
  \bibinfo{volume}{151} (\bibinfo{year}{2021}), \bibinfo{pages}{125--137}.
\newblock


\end{thebibliography}

\end{document}